\documentclass{article}

\usepackage{arxiv}

\usepackage[utf8]{inputenc} 
\usepackage[T1]{fontenc}    
\usepackage{hyperref}       
\usepackage{url}            
\usepackage{booktabs}       
\usepackage{amsfonts}       
\usepackage{nicefrac}       
\usepackage{microtype}      
\usepackage{graphicx}
\usepackage{doi}
\usepackage{algorithmic}
\usepackage{bm}
\usepackage{amsmath}

\title{Use of Tensor-Train Decompositions with a Discrete Velocity Boltzmann Solver}

\author{ \href{https://orcid.org/0000-0002-1434-7166}{\includegraphics[scale=0.06]{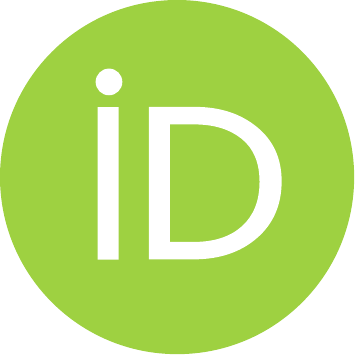}\hspace{1mm}Georgii Oblapenko}\thanks{
Preprint submitted to SIAM Journal on Scientific Computing.} \\
	German Aerospace Center (DLR), \\
	Bunsenstrasse 10, G\"{o}ttingen, Germany \\
	\texttt{georgii.oblapenko@dlr.de}
}

\renewcommand{\shorttitle}{Use of Tensor-Train Decompositions with a Discrete Velocity Boltzmann Solver}

\newcommand{\vect}[1]{\boldsymbol{\mathbf{#1}}}
\hypersetup{
pdftitle={Use of Tensor-Train Decompositions with a Discrete Velocity Boltzmann Solver},
pdfsubject={q-bio.NC, q-bio.QM},
pdfauthor={Georgii Oblapenko},
pdfkeywords={ Boltzmann equation, discrete velocity method, tensor decomposition, tensor train, rarefied gas, low-rank approximation},
}

\begin{document}
\maketitle

\begin{abstract}
	In the present work, the Tensor-Train decomposition algorithm is applied to reduce the memory footprint of a stochastic discrete velocity solver for rarefied gas dynamics simulation. An energy-conserving modification to the algorithm is proposed, along with an interleaved collision/convection routine which allows for easy application of higher-order convection schemes. The performance of the developed algorithm is analyzed for several 0- and 1-dimensional model problems in terms of solution error and reduction in memory use requirements.
\end{abstract}


\section{Introduction}
The motion of gas or fluid can be successfully described by a set of partial different equations under a wide range of conditions of interest. However, in rarefied regimes, when the ratio of the molecular mean free path to the characteristic length scale becomes large, the continuum approaches no longer apply, and instead, the full Boltzmann equation describing the evolution of the velocity distribution function due to advection, external forces, and collisions, has to be solved. It is a complicated integro-differential equation, with a 6-dimensional integral (the collision operator) appearing on the right-hand side. Numerous approaches have been developed over the years to tackle the equation. These include stochastic approaches, such as the Direct Simulation Monte Carlo (DSMC) method~\cite{bird1994molecular}, and deterministic approaches, such as discrete velocity methods~\cite{platkowski1988discrete} and spectral methods~\cite{filbet2006solving,gamba2017fast}.

The family of discrete velocity methods (DVM), which is the main focus of the present work, operates by discretizing the velocity distribution function on a grid in velocity space, and obtains a system of partial differential equations for the values of the distribution function at each grid node. A significant advantage offered by discrete velocity methods is the noticeable reduction in noise compared to particle-based methods, which is needed to better understand the small time-scale dynamics of complex rarefied flows, especially in unsteady scenarios~\cite{hara2012one, chan2022enabling, chan2023grid}.

Within the discrete velocity framework, the complex collision operator is often replaced with a simple relaxation term~\cite{mieussens2000discrete,dimarco2014numerical}. Such relaxation terms include the Bhatnagar–Gross–Krook (BGK) model~\cite{bgk} and its extensions, the ellipsoidal-statistical BGK model (ES-BGK)~\cite{holway1966new} and the Shakhov model (S-model)~\cite{shakhov1968generalization}. However, the correct incorporation of complex collisional physical phenomena, such as chemical reactions, transitions of internal energy, and complicated scattering laws within the framework of these model equations is still an open question and a topic of active research~\cite{bisi2021kinetic, mathiaud2022bgk, haack2023numerical}. Despite these drawbacks of the linear models, they offer advantages in terms of being scalable to dense regimes~\cite{xu2010unified, pfeiffer2022exponential}.

Another subfamily of the discrete velocity methods does not resort to substituting the collision operator with a simplified model relaxation term, and instead models the full collision process, utilizing remapping procedures to re-distribute post-collisional mass onto the discrete velocity grid~\cite{tcheremissine1998conservative,morris2011monte}. These approaches have been shown to be capable of simulating more complex collisional processes, such as internal energy relaxation~\cite{tcheremissine2012method,clarke2018low}, chemical reactions~\cite{poondla2022modeling}, and ionization~\cite{oblapenko2021velocity,oblapenko2021modeling}, and can make use of state-specific cross-sections and anisotropic scattering laws.

However, the necessity of selecting a sufficiently accurate velocity grid remains a significant issue in the application of discrete velocity methods~\cite{chan2023grid}. The associated memory cost can also be a significant detriment. An example of an approach that aims to remedy the issue is the velocity-space octree grid developed in~\cite{kolobov2007unified}. Another family of approaches, which has been receiving increased attention, is the use of (dynamic) low-rank methods, which project the equations of interest (not necessarily restricted to gas dynamic equations) onto a lower-dimensional manifold. The area has seen significant development, with recent work focusing on improving the adaptivity via efficient error estimates~\cite{hauck2022predictor}, ensuring conservation of mass/momentum/energy~\cite{einkemmer2021mass,guo2022local}, and applying low-rank methods to the continuum regime in the presence of discontinuities~\cite{einkemmer2021efficient}.
Such low-rank methods are of significant interest not only due to the reduced computational and memory cost, but also due to their potential for use in machine learning-based Boltzmann solvers~\cite{xiao2021using,holloway2021acceleration}, where the lower input dimensionality can simplify the reduced-order model architecture and reduce its computational cost. They can also be of use as a storage reduction technique for simulation restarts~\cite{chen2021unsupervised}.

In~\cite{kornev2020tensorized,chikitkin2021numerical}, the Tensor Train representation~\cite{oseledets2011tensor} was used to obtain a reduced-order representation of the velocity distribution function, with the approach differing from most other works in that the reduced-order representation of the velocity distribution function was stored independently for each cell on the physical space grid, which allows for the use of unstructured grids, whereas in previous studies~\cite{khoromskij2007structured,dolgov2014low,einkemmer2021mass}, the decomposition was applied to the multi-dimensional Cartesian product of the (structured) physical and velocity domains. A similar approach was utilized later on in \cite{allmann2022parallel} for the Vlasov--Maxwell equations, with low-rank decomposition applied only in velocity space and not the full 6-dimensional physical and phase space.

However, the algorithm, as developed in~\cite{kornev2020tensorized,chikitkin2021numerical}, has several deficiencies: 1) as it operates directly on the reduced-order representation of the distribution function, it is restricted to simple model collisional terms 2) the tensor decomposition might introduce errors in the conserved quantities (number density, momentum, energy) 3) due to constraints on the tensor representation, the algorithm cannot compute the exact numerical flux needed for convection, and issues arise with the implicit time-stepping scheme. Finally, a more comprehensive numerical analysis of the errors introduced by the application of tensor decomposition was also not carried out. Other low-rank approaches also usually consider either a model collision term, or collisionless plasma flows; as such, their applicability to simulations of collisional flows with more complex Boltzmann collision terms has not been thoroughly investigated. In a recent paper~\cite{hu2022adaptive}, the fast spectral method~\cite{mouhot2006fast} was used to compute the full collision operator; however, the work relied on a coupled physical/phase space low-rank decomposition, and did not consider the impact of violation of conservation laws.
 In addition, the evaluation of the full collision term (even with a fast spectral method) is computationally expensive; as such, one might resort to the use of stochastic approximations, which introduce noise into the simulation, and it is not clear how the fidelity of low-rank representations is affected by the presence of numerical noise in the velocity distribution function.

Thus, in the present work, a variation on the tensor decomposition idea is developed. Instead of a model equation, the full Boltzmann collision operator is computed, using the Quasi-Particle Simulation (QUIPS) discrete velocity method~\cite{clarke2018low}. Whilst this requires reconstructing the full velocity distribution function, an interlaced collision-convection algorithm is proposed, which not only allows to maintain reduced memory usage (compared to a DVM solution without the use of tensor decomposition), but also allows for correction of errors in the number density and energy introduced by the tensor decomposition.

The paper is organized as follows. In~\ref{sec:alg}, a description of the discrete velocity-based quasi-particle simulation method is given. Then in~\ref{sec:tt-alg} the reduced-order representation of the velocity distribution function via the Tensor-Train decomposition and the associated modifications of the discrete velocity-based kinetic solver are presented. Numerical results and their analysis  are presented in~\ref{sec:numerical} for model 0-dimensional and 1-dimensional flows. Finally, conclusions and discussion of future work follow in~\ref{sec:conclusions}.

\section{Discrete velocity quasi-particle method}
\label{sec:alg}
\subsection{Discretization}
Let us denote by $f(\vect{x}, \vect{v}, t)$ (and hereafter refer to as the velocity distribution function) the density of the mathematical expectation of the number of particles in an element of phase space $(\mathbf{x}, \mathbf{x}+d\mathbf{x})$, $(\vect{v}, \vect{v} + d\vect{v})$ at time $t$, where $\vect{x}$ is the vector corresponding to the spatial coordinate, and $\vect{v}$ is the vector corresponding to the velocity of the particles.

The Boltzmann equation for a monoatomic gas (without internal degrees of freedom) without external forces reads as follows (omitting the spatial coordinate $\vect{x}$ and time $t$ for brevity):
\begin{equation}
    \frac{\partial f(\vect{v}_1)}{\partial t} + \vect{v}_1 \cdot \nabla_{{\vect{x}}} f = \int \left[f({\vect{v}_1}')f({\vect{v}_2}') - f({\vect{v}_1})f(\vect{v}_2)\right]g{\sigma}d^2\Omega d\vect{c}_2.\label{eq:Boltzmann}
\end{equation}
Here  $\vect{v}_2$ is the velocity of the collision partner, $g=\left|{\vect{v}_1} - {\vect{v}_2}\right|$ is the magnitude of the relative collision velocity, $\sigma = \sigma(g)$ is the collision cross-section, $d^2 \Omega$ is the solid angle into which the post-collision relative velocity vector is scattered, and primed variables denote the post-collisional quantities.

Within the framework of the discrete velocity method,  the velocity distribution function is assumed to be defined only at a discrete set of points $\mathcal{D} \in \mathbb{R}^3$. While the choice of the discrete velocities $\mathcal{D}$ is, generally speaking, arbitrary, for practical purposes $\mathcal{D}$ is often assumed to be a simple Cartesian product of velocity grids defined independently for each of the three components:
\begin{equation}
    \mathcal{D} = \mathcal{D}_x \times \mathcal{D}_y \times \mathcal{D}_z.
\end{equation}
In the present work, only uniform grids with equal numbers of points $N_v$ and equal velocity spacing $\Delta v$ in each velocity direction are considered:
\begin{equation}
    \mathcal{D} = \left\{\vect{v}_{ijk} \right\}_{i,j,k = 1}^{i,j,k = N_v},
\end{equation}
where
\begin{equation}
    \vect{v}_{ijk} = \left(v_{x,min} + i \Delta v, v_{y,min} + j \Delta v, v_{z,min} + k \Delta v \right) = \left(v_{x,i}, v_{y,j}, v_{z,k} \right).
\end{equation}
Here $v_{x,min}$, $v_{y,min}$, $v_{z,min}$ are the minimum values of the x-, y-, and z-velocities at which the distribution function is defined.

Then, the discrete velocity distribution function can be defined as follows:
\begin{equation}
    \hat{f}_{ijk}(\vect{x}, t) = (\Delta v)^3 f(\vect{x}, \vect{v}_{ijk}, t),\:i,j,k=1,\ldots,N_v,
\end{equation}
and the Boltzmann equation~(\ref{eq:Boltzmann}) re-written in discrete form:
\begin{equation}
    \frac{\partial  \hat{f}_{i_1j_1k_1}}{\partial t} + \vect{v}_{i_1j_1k_1} \cdot \nabla_{{\vect{x}}} \hat{f}_{i_1j_1k_1} = \int \sum_{i_2 j_2 k_2} \left[\hat{f}_{i_1'j_1'k_1'} \hat{f}_{i_2'j_2'k_2'} - \hat{f}_{i_1j_1k_1}\hat{f}_{i_2 j_2 k_2}\right]g{\sigma}d^2\Omega. \label{eq:boltzmann_discrete}
\end{equation}

Of course, even in such a form, the integration over the possible scattering angles $\Omega$ remains to be defined, since the post-collision velocities need to lie on the grid points. Thus, given the initial coordinates (in velocity space) $i_1,j_1,k_1$ and $i_2, j_2, k_2$ of the colliding velocities, an algorithm is required to compute the indices of the post-collision velocities $i_1',j_1',k_1'$ and $i_2',j_2',k_2'$.

\subsection{Collisions}
A conceptually simple way to compute collisions is to restrict oneself to post-collision velocities which lie directly on grid nodes~\cite{goldstein1989investigations}. However, such a procedure not only becomes increasingly complicated as internal energy transitions are taken into account (as collisions no longer conserve translational energy and the magnitude of the post-collision velocity vector is altered), but also suffers from a slow convergence rate~\cite{dimarco2014numerical}.

An alternative approach is the use of remapping (also referred to in literature as ``interpolation'') procedures~\cite{tcheremissine1998conservative,tcheremissine2005direct,varghese2007arbitrary,morris2008improvement}, in which the post-collision velocity is not restricted to the discrete velocity grid, but after a collision is performed, the remapping/interpolation procedure is used to update the velocity distribution function at the neighbouring grid points, ensuring mass, momentum, and energy conservation. Thus, the first step collision process is performed akin to DSMC --- a post-collision velocity vector is computed based on the specific cross-section used, and only after the collision has been performed, the remapping procedure is applied. In the present work, the remapping procedure developed in~\cite{varghese2007arbitrary,morris2008improvement,morris2011monte}, and later extended to non-uniform velocity grids~\cite{clarke2018low}, is used. It allows for easy incorporation of internal energy transitions~\cite{clarke2018low}, chemical reactions~\cite{poondla2020modeling,poondla2022modeling} and anisotropic scattering laws~\cite{oblapenko2021modeling}.
Here, a brief overview of the remapping procedure is given for points lying inside the domain bounded by $\mathcal{D}$; for a more detailed description, including the treatment of velocities that fall outside of the domain bounded by $\mathcal{D}$ and generalization to non-uniform grids, the reader is referred to~\cite{clarke2018low}.
Given a post-collision velocity $\mathbf{c}' = \left(c_x', c_y', c_z'\right)$ and a number density $n_0$ at that velocity,  the indices $i, j, k$ of the closest velocity grid point are computed:
\begin{equation}
    i = \mathrm{int}\left(\frac{c_x' - v_{x,min}}{\Delta v}\right),
    \:
    j = \mathrm{int}\left(\frac{c_y' - v_{y,min}}{\Delta v}\right),
    \:
    k = \mathrm{int}\left(\frac{c_z' - v_{z,min}}{\Delta v}\right).
\end{equation}
Here $\mathrm{int}$ denotes rounding to the nearest integer. Then, the velocity relative to the found grid point is calculated: $\Delta \mathbf{c}' = \mathbf{c}' - \mathbf{v}_{ijk} = \left(\Delta c_x', \Delta c_y', \Delta c_z'\right)$. Next, the indices for the so-called ``internal'' remapping points are computed:
\begin{equation}
    i_{int} = i + \mathrm{sign}(\Delta c_x'),\:
    j_{int} = j + \mathrm{sign}(\Delta c_y'),\:
    k_{int} = k + \mathrm{sign}(\Delta c_z'),
\end{equation}
as well as indices for the ``external'' remapping points:
\begin{equation}
    i_{ext} = i - \mathrm{sign}(\Delta c_x'),\:
    j_{ext} = j - \mathrm{sign}(\Delta c_y'),\:
    k_{ext} = k - \mathrm{sign}(\Delta c_z').
\end{equation}
After these indices have been computed, the number density is redistributed amongst 7 points defined by these indices:
\begin{equation}
    (i,j,k),\:(i_{int},j,k),\:(i,j_{int},k),\:(i,j,k_{int}),\:(i_{ext},j,k),\:(i,j_{ext},k),\:(i,j,k_{ext}).
\end{equation}
As only 5 conservation laws (one for mass, three for momentum, and one for energy) need to be satisfied, an additional constraint is imposed: the number density redistributed to the external points is assumed to be split equally. However, other constraints can be imposed, for example, conservation of directional second-order moments (and not just the full energy).
The number density to be distributed to each of the points is found by solving the following linear system:
\begin{equation}
    \begin{bmatrix}
1 & 1 & 1 & 1 & 3 \\
0 & 1 & 0 & 0 & -1 \\
0 & 0 & 1 & 0 & -1 \\
0 & 0 & 0 & 1 & -1 \\
0 & 1 & 1 & 1 & 3
\end{bmatrix}
\begin{bmatrix}
 n_{or} \\ n_{int,x} \\ n_{int,y} \\n_{int,z} \\ n_{ext}
\end{bmatrix}
=
n_0 \begin{bmatrix}
1 \\ \Delta c_x' / \Delta v \\ \Delta c_y' / \Delta v \\ \Delta c_z' / \Delta v \\
\left((\Delta c_x')^2 + (\Delta c_y')^2 + (\Delta c_z')^2\right) / (\Delta v)^2.
\end{bmatrix}
\end{equation}
Here, $n_{or}$ is the number density to be added to the point $(i,j,k)$, $n_{int,x}$, $n_{int,y}$, $n_{int,z}$ are the number densities to be added to the internal remapping points, and $n_{ext}$ is the number density to be added to each of the external points.

Finally, in the present work, a stochastic scheme is used to evaluate the summation that appears in the right-hand side of Eqn.~(\ref{eq:boltzmann_discrete}), as a direct summation leads to an extremely high computational cost of order $~\mathcal{O}\left(N_v^6\right)$. Instead, importance sampling is performed to select two pre-collision velocities on the grid, with the probability of selection being proportional to the value of $\left|\hat{f} \right|$ at those grid nodes. The value of the distribution function at those nodes is depleted by a quantity $\delta \hat{f}$, and only this (small) quantity $\delta \hat{f}$ is collided and subsequently remapped.
The value of $\delta \hat{f}$ is computed as
\begin{equation}
    \delta \hat{f} = (\Delta v)^3 \frac{\Delta t}{2 N_c} (n - 2n_-)^2 g\sigma(g) \mathrm{sign}(\hat{f}_1 \hat{f}_2).
\end{equation}
Here $\Delta t$ is the timestep, $n$ is the number density in the current computational cell, $n_-$ is the negative number density in the current computational cell (see below for a discussion on the presence of negative values of the distribution function), $\hat{f}_1$ and $\hat{f}_1$ are the values of the velocity distribution function at the velocity grid nodes chosen for collision, and $N_c$ is the number of collisions performed. The quantity $N_c$ is computed as
\begin{equation}
    N_c = \left\lfloor \frac{\Delta t}{2C_{RMS}} \left(\frac{n - 2n_-}{n}\right)^2 \frac{1}{(\Delta v)^3} \left(\frac{2kT}{m}\right)^2{\exp\left[\left(\frac{m v_{max}^2}{2kT}\right)^{\alpha}
    \right]}{n_r \sigma_r} + \mathcal{R}\right\rfloor.\label{eq:ncoll}
\end{equation}
Here $\sigma_r$ is a reference cross-section evaluated at a reference temperature $T_r$, $n_r$ is a reference number density, $v_{max}$ is the maximum speed possible on the velocity grid, $C_{RMS}$ is a user-adjustable parameter that governs the amount of numerical noise in the simulation due to the computation of collisions, and $\alpha$ is a parameter that governs how strongly the changes in the temperature and/or velocity grid extent affect the number of collisions performed. $R$ is a random number uniformly distributed between 0 and 1, and the brackets denote rounding down to the nearest integer.

As in~\cite{poondla2022modeling}, the value of $N_c$ is computed taking into account the minimum value of the equilibrium distribution function on the defined grid (via the presence of the $\exp\left[\left(\frac{m v_{max}^2}{2kT}\right)^{\alpha}\right]$ factor). However, in~\cite{poondla2022modeling}, $\alpha$ was taken to be 1. Assuming a fixed velocity grid with a $v_{max}=\sqrt{3}\times 3.5 \sqrt{{2kT_{ref}}/{m}}$ (which corresponds to the maximum speed on a cubic grid with an extent of $\pm 3.5 \sqrt{{2kT_{ref}}/{m}}$ (here $T_{ref}$ is some fixed reference temperature), a 2-fold increase in $T$ (compared to the reference temperature $T_{ref}$) leads to an approximately 2$\times 10^7$-fold reduction in $N_c$. Conversely, a 2-fold decrease in temperature compared to $T_{ref}$ leads to an approximately $2\times 10^{15}$ increase in the number of collisions to be performed. Obviously such extreme changes in $N_c$ are not feasible from a computational point of view. In~\cite{poondla2022modeling}, an adaptive grid cut-off procedure was proposed to reduce the velocity space extent when the temperature decreases, so as to avoid the issue of the collision number blow-up. Another option is to implement a temperature-dependent $C_{RMS}$ value. In the present work, the $\alpha$ factor is introduced, so that the impact of temperature variations is not so drastic, especially if the grid extent is kept fixed. With $\alpha=0.5$ (the value used in the present study), a 2-fold increase in temperature leads to just an approximately 1.5-fold reduction in the number of collisions being performed, whereas a 2-fold decrease in temperature leads to a 3-fold increase in $N_c$. 

Higher values of $C_{RMS}$ lead to higher levels of noise in the velocity distribution function (as the depletion quantity $\delta \hat{f}$ becomes large), but a lower computational cost (as fewer collisions are performed). One of the drawbacks of the described algorithm is the presence of negative values of the distribution function. These occur due to 1) the remapping procedure 2) the depletion quantity $\delta \hat{f}$ being larger than the value of the distribution function at the node being depleted. At extremely high values of $C_{RMS}$ simulations can become unstable, as the large percentage of negative number density leads to an increase in collisions simulated (as the quantity $(n-2n_-) / n$ grows), which in turn leads to even more negative number density being produced as a result of many depletion events, which leads to a runaway effect. However, despite this shortcoming, the described algorithm has been shown to be capable of modelling of various rarefied gas flows with a high level of accuracy, and the negative number densities can be handled in a consistent manner by appropriately defining the depletion quantity and number of collisions.
It should be noted that different definitions of $N_c$ are possible, as long as the quantity $\delta \hat{f}$ is defined consistently with $N_c$ and the negative values of the distribution function are accounted for.

\subsection{Convection}
\label{sec:convection}
The last step requiring description is the computation of the advection of the velocity distribution function. In the present work, a second-order finite volume scheme with a min-mod flux limiter is utilized on a Cartesian spatial grid~\cite{baranger2019numerical}. At boundaries, the scheme becomes a first-order one, although higher-order monotonicity-preserving discrete-velocity schemes at the boundary can also be utilized (however, a re-formulation of the flux limiting procedure to a slope limiting procedure is required~\cite{baranger2019numerical}). Only 1-dimensional problems (with a spatial coordinate in the x-direction) are considered, and a uniform spatial grid is assumed for simplicity, however, the scheme is easily extended to 2- and 3-dimensional Cartesian physical grids.

Denoting by $\hat{f}_{ijk}^{l,n}$ the cell-average of the distribution function $\hat{f}_{ijk}$ in cell $n$ at timestep $l$, a first-order explicit timestepping scheme reads:
\begin{equation}
    \frac{\hat{f}_{ijk}^{l+1,n} - \hat{f}_{ijk}^{l,n}}{\Delta t} = \mathcal{Q}(\hat{f}_{ijk}^{l,n}) - \frac{\mathcal{F}^{l,n+\frac{1}{2}}_{ijk} - \mathcal{F}^{l,n-\frac{1}{2}}_{ijk}}{\Delta x},
\end{equation}
where $\mathcal{Q}(\hat{f}_{ijk}^{l,n})$ is the approximate collision integral computed using the Monte-Carlo procedure outline above, $\mathcal{F}^{l,n+\frac{1}{2}}_{ijk}$ is the numerical flux at the interface between cells $n$ and $n+1$, and $\Delta x$ is the spatial grid spacing.

The second-order flux is computed as
\begin{equation}
    \mathcal{F}^{l,n+\frac{1}{2}}_{ijk} = v_{x,i}^{+}\hat{f}_{ijk}^{l,n} + v_{x,i}^{-}\hat{f}_{ijk}^{l,n+1} + |v_{x,i}|\Phi^{l,n+\frac{1}{2}}_{ijk},
\end{equation}
where $v_{x,i}^{\pm} = (v_{x,i} \pm |v_{x,i}|)/2$, and $\Phi^{l,n+\frac{1}{2}}_{ijk}$ is defined as
\begin{equation}
    \Phi^{l,n+\frac{1}{2}}_{ijk} = \mathrm{minmod}\left(
    \Delta \hat{f}_{ijk}^{l,n-\frac{1}{2}},
    \Delta \hat{f}_{ijk}^{l,n+\frac{1}{2}},
    \Delta \hat{f}_{ijk}^{l,n+\frac{3}{2}}\right),
\end{equation}
where $\Delta \hat{f}_{ijk}^{l,n+\frac{1}{2}} = \hat{f}_{ijk}^{l,n+1} - \hat{f}_{ijk}^{l,n}$, and the minmod function is given by
\begin{equation}
    \mathrm{minmod}(x,y,z) = \begin{cases}
    \mathrm{sign}(x)\mathrm{min}(|x|, |y|, |z|),& \text{if } \mathrm{sign}(x)=\mathrm{sign}(y)=\mathrm{sign}(z)\\
    0,              & \text{otherwise}
\end{cases}.
\end{equation}
Finally, the boundary conditions are imposed via the use of ghost cells. Assuming that the actual domain is discretized by cells with indices ranging from $1$ to $N_x$, 4 additional ghost cells are used (2 on each side), with indices -1, 0, and $N_x + 1$, $N_x + 2$.
In the present work, only fully diffusive boundary conditions are considered (the reflected velocity distribution function is a half-Maxwellian at a pre-defined wall temperature and velocity), and thus, one needs to compute the incoming and outgoing fluxes at the boundaries in order to ensure mass conservation.
For ghost cells with indices -1 and $N_x + 2$, a zero-order extrapolation is used:
\begin{equation}
    \hat{f}_{ijk}^{l,-1} = \hat{f}_{ijk}^{l,0};\:\hat{f}_{ijk}^{l,N_x+2} = \hat{f}_{ijk}^{l,N_x + 1}.
\end{equation}
Thus, only the values of $\hat{f}_{ijk}^{l,0}$ and $\hat{f}_{ijk}^{l,N_x + 1}$ remain to be defined.
For outgoing velocities ($v_x < 0$ at the left boundary and $v_x > 0$ at the right boundary), a zero-order extrapolation is used:
\begin{equation}
    \hat{f}_{ijk}^{l,0} = \hat{f}_{ijk}^{l,1}\quad\forall i:\:v_{x,i} < 0;\quad \hat{f}_{ijk}^{l,N_x + 1} = \hat{f}_{ijk}^{l,N_x}\quad\forall i:\:v_{x,i} > 0.
\end{equation}
For incoming velocities ($v_x > 0$ at the left boundary and $v_x < 0$ at the right boundary), the velocity distribution function is computed as
\begin{equation}
    \hat{f}_{ijk}^{l,0} = \frac{\sum_{jk}\left|v_{x,i}^{-}\right|\hat{f}_{ijk}^{l,1}}{\sum_{ijk}v_{x,i}^{+}\left(\Delta v\right)^3 \mathcal{M}\left(v_{y,lw},v_{z,lw},T_{lw}\right)},
\end{equation}
\begin{equation}
    \hat{f}_{ijk}^{l,N_x+1} = \frac{\sum_{jk}v_{x,i}^{+}\hat{f}_{ijk}^{l,N_x}}{\sum_{ijk}\left|v_{x,i}^{-}\right|\left(\Delta v\right)^3 \mathcal{M}\left(v_{y,rw},v_{z,rw},T_{rw}\right)}.
\end{equation}
Here $\mathcal{M}$ denotes a Maxwellian distribution function evaluated on the discrete grid, $v_{y,lw}$ and $v_{z,lw}$ are the y- and z-velocities of the left wall, correspondingly, $T_{lw}$ is the temperature of the left wall, $v_{y,rw}$ and $v_{z,rw}$ are the y- and z-velocities of the right wall, correspondingly, and $T_{lw}$ is the temperature of the right wall.

Thus, all the necessary constituent parts of the algorithm have been touched upon: 1) the discretization, 2) the collision selection procedure, 3) the remapping procedure, and 4) the convection scheme. In the next section, the tensor-decomposition version of the quasi-particle simulation method will be introduced and elaborated upon.

\section{Tensor decomposition-based quasi-particle simulation method}
\label{sec:tt-alg}
\subsection{Algorithm description}
In the present work, the Tensor Train reduced-order representation format~\cite{oseledets2011tensor,zniyed2020tt} is utilized, which approximates a tensor $\mathcal{A}_{i_1,\ldots,i_D} \in \mathbb{R}^{\left(n^D\right)}$ (assuming that $1 \leq i_k \leq n$, $k=1,\ldots,D$) in the following form:
\begin{equation}
    \mathcal{A}_{i_1,\ldots,i_D} \approx \mathcal{G}_1(i_1)\mathcal{G}_2(i_2)\ldots\mathcal{G}_D(i_D),\label{eq:tt-decomp}
\end{equation}
where $\mathcal{G}_1(i_1)$ is a row vector, $\mathcal{G}_D(i_D)$ is a column vector, and $\mathcal{G}_k(i_k),\:k=2,\ldots,D-1$ are matrices. Whilst the machine memory  required for storing $\mathcal{A}$ is $\mathcal{O}\left(n^{D}\right)$, the memory required for storing the representation~(\ref{eq:tt-decomp}) is  $\mathcal{O}\left(nDr^2\right)$, where $r$ is the upper bound of the ranks $r_k$ of $\mathcal{G}_k(i_k)$. In the current work, the main focus is on the tensor decomposition of the velocity distribution function (which is a tensor with dimension $D=3$), and thus the expression above can be simplified, also assuming that all ranks $r_k$ are equal:
\begin{equation}
    \hat{f}_{ijk} \approx \mathcal{G}_1(i) \mathcal{G}_2(j) \mathcal{G}_3(k),
\end{equation}
where $\mathcal{G}_1(i) \in \mathbb{R}^{1 \times r}$, $\mathcal{G}_2(j) \in \mathbb{R}^{r \times r}$, $\mathcal{G}_3(k) \in \mathbb{R}^{r \times 1}$.

The computation of decomposition~(\ref{eq:tt-decomp}) that minimizes the quantity
\begin{equation}
    \epsilon_{ijk,TT} = \left|\left| \hat{f}_{ijk} - \hat{f}_{ijk,TT} \right|\right| \label{eq:tt-error}
\end{equation}
requires $\mathcal{O}\left(nDr^3\right)$ operations. Here $\left|\left| M \right|\right|$ is the Frobenius norm of a matrix $M$, and $\hat{f}_{ijk,TT}$ is the reduced-order tensor decomposition-based representation of the VDF $\hat{f}_{ijk}$ (further this reduced-order representation of the VDF will also be referred to as ``compressed'').

\subsection{Energy-conservative modified collision algorithm}
In case the tensorized representation cannot represent the VDF without introducing some error ($\epsilon_{ijk,TT} \ne 0$), mass, momentum, and energy will not be conserved. Mass conservation is the easiest to fix: before compressing $\hat{f}_{ijk}$ in a cell $n$, the number density in that cell is computed and stored; when the VDF is reconstructed during the next timestep, it is re-normalized so that the previously stored number density is recovered.
In order to ensure energy conservation, the following procedure is proposed, based on a similar idea developed for variable-weight DSMC simulations~\cite{boyd1996conservative}. Before the distribution function $\hat{f}^{l,n}$ in spatial cell $n$ at timestep $l$ is ``compressed'' (converted to a tensor decomposition-based representation), the energy $E^{l,n}$ in the cell is computed and stored. After the VDF is reconstructed from the reduced-order representation, the post-reconstruction energy is computed $E^{l,n\ast}$, as well as the change in energy due to the tensor decomposition operation: $\Delta E = E^{l,n} - E^{l,n\ast}$.
Then, before the computation of the collision step, an average per-collision energy adjustment is defined as $\delta E = \Delta E / N_c$. This quantity $\delta E$ is added to the post-collision translation energy in order to produce a post-collision scattering velocity with magnitude $g'$ computed through the energy conservation law:
\begin{equation}
    \frac{mg^2}{2} = \frac{m(g')^2}{2} + \delta E.\label{eq:de-fix}
\end{equation}
So if $\Delta E$ is greater than 0, the energy in the collisions will be decreased, and if $\Delta E < 0$, the energy in the collisions will be increased. Since Eqn.~\ref{eq:de-fix} may not always permit a solution $g'$, the quantity $\delta E$ is recomputed after each collision via keep track of the remaining number of collisions and the fraction of the already re-distributed energy $\Delta E$.

\subsection{Modified advection algorithm}
Finally, advection has to be discussed. Since a number of mathematical operations can be performed on the compressed VDF representation directly, one can continue using the usual operator splitting approach: first, compute collisions in all spatial grid cells (decompressing and re-compressing the VDF), and then compute the changes in the velocity distribution function in all spatial grid cells due to convection by operating directly on the Tensor Train representation.
However, as pointed out in~\cite{kornev2020tensorized}, the computation of the numerical flux $\mathcal{F}_{ijk}^{l,n+\frac{1}{2}}$ is not exact even in the first-order case in case of physical grid cells with normals that are not aligned with the coordinate axes, and the non-smooth absolute value function is needed for the flux computation. As such, the numerical flux is only approximated. In addition, some of the operations on the reduced-order tensor representations (such as addition) increase the rank of the compressed representation, requiring periodic re-compression, and thus introducing additional error in the moments of the distribution function, including the conserved quantities (mass, momentum, energy).
The use of higher-order schemes and the necessity of using flux or slope limiting also poses a question of the applicability of the approach of computing convection via direct manipulation of the compressed VDF representation. In the present work, a different approach is proposed.
From the description of the convection given in~\ref{sec:convection}, it can be seen that in order to compute the new value of the velocity distribution function in cell $n$, one needs the values of the velocity distribution function in cells $n$, $n-1$ and $n-2$ in order to compute $\mathcal{F}^{l,n-\frac{1}{2}}_{ijk}$, and in cells $n$, $n+1$ and $n+2$ in order to compute $\mathcal{F}^{l,n+\frac{1}{2}}_{ijk}$.

Let us assume that at the start of timestep $l$, one only has the compressed representations of $\hat{f}_{ijk}^{l,n}$. First, the values of $\hat{f}_{ijk}^{l,n}$ are reconstructed in cells $n=1,2,3$. Then the approximate collision operator $\mathcal{Q}(\hat{f}_{ijk}^{l,n})$ is computed in these cells, and the values of the VDF in the ghost cells $n=-1,0$ are also computed. This provides sufficient information to compute $\hat{f}_{ijk}^{l+1,1}$. Whilst not enough information is available to compute $\hat{f}_{ijk}^{l+1,2}$ (as $\hat{f}_{ijk}^{l,4}$ is required for the computation of $\mathcal{F}^{l,2+\frac{1}{2}}_{ijk}$), sufficient information to compute the left-side flux $\mathcal{F}^{l,2-\frac{1}{2}}_{ijk}$ is available. The VDF in cell $n=-1$ (as it is no longer required) is then compressed, the VDF in cell $4$ reconstructed and the quantity $\mathcal{Q}(\hat{f}_{ijk}^{l,4})$ computed. Now sufficient information is available to complete the update of the VDF in cell $2$, as the flux $\mathcal{F}^{l,2+\frac{1}{2}}_{ijk}$ can be computed; the flux $\mathcal{F}^{l,3-\frac{1}{2}}_{ijk}$ is also computed and stored. The VDF in cell $n=0$ is then compressed and the VDF in cell $5$ reconstructed. The process (compute collisions in cell $n$, compute missing right-hand side flux in cell $n-2$, update the VDF in cell $n-2$, and compute the left-hand side flux in cell $n-1$) is repeated until the right-hand side boundary is reached. Note: the compression/reconstruction of the VDF in the ghost cells does not actually have to be performed, as its value is computed from the adjacent cells in the domain; however, the algorithm description presented assumes such a procedure for the sake of consistency.

Such an approach allows to use any method for flux computation and flux/slope limiting, without introducing any numerical error due to operating on the compressed velocity distribution function representations. The storage requirements for such an interleaved/convection algorithm are thus 
\begin{equation}
    N_x \times \mathrm{mem}(TT(VDF)) + 6 \times \mathrm{mem}(VDF),
\end{equation}
where $\mathrm{mem}(x)$ is the machine memory used by some quantity $x$, and $TT(VDF)$ denotes the Tensor Train representation of the VDF. The factor of 6 comes from the fact that one needs to have access to the full distribution function in 5 grid cells, as well as store the flux.
In the case of a 2-dimensional Cartesian physical grid with $N_x \times N_y$ cells, a similar reasoning can be applied, but it does require the reconstruction of the VDF across a ``slice'' of the domain, and therefore, the storage requirements become
\begin{equation}
    N_x N_y\times \mathrm{mem}(TT(VDF)) + 6 \times \mathrm{min}(N_x, N_y)\mathrm{mem}(VDF).
\end{equation}
In the case of a 3-dimensional Cartesian grid with $N_x \times N_y \times N_z$ cells, the storage requirements become (assuming that the propagation direction is chosen in such a way so as to minimize memory use)
\begin{equation}
    N_x N_y N_z \times \mathrm{mem}(TT(VDF)) + 6 \times \mathrm{min}(N_xN_y,N_xN_z,N_yN_z)\mathrm{mem}(VDF).
\end{equation}

\section{Numerical results}
\label{sec:numerical}
To validate and study the performance of the developed algorithm, several model problems are simulated: the BKW relaxation, a Couette flow, and a Fourier flow.
The solver was written in the Julia programming language~\cite{bezanson2017julia} and makes use of the TensorToolbox.jl library~\cite{kressner2017recompression,ttjl} for the tensor decomposition.

\subsection{BKW relaxation}
The Bobylev-Krook-Wu~\cite{bobylev1976, krook1977} relaxation problem is an unsteady spatially homogeneous model problem, widely used for verification of Boltzmann solvers, as it provides an analytic expression for the evolution of the velocity distribution function in time, assuming only that the interaction potential is pseudo-Maxwell; that is, $\sigma(g) = C_{\sigma} g^{-1}$, where $C_{\sigma}$ is a constant. The $C_{\sigma}$ constant can be computed from the standard Variable Hard Sphere (VHS) interaction model~\cite{bird1994molecular} parameters:

\begin{equation}
    C_{\sigma} = 2 \pi d^2 \frac{\left(2 k T_{VHS}  / m_r \right)^{\omega - 1/2}}{ \Gamma(5/2 - \omega)},
\end{equation}
where $d$ is the reference VHS diameter at a given reference temperature $T_{VHS}$, $m_r=m/2$ is the collision-reduced mass, and $\omega$ is the viscosity exponent, equal to 1 in the case of Maxwell molecules.

As the problem considered is purely abstract,  the number density $n$, temperature $T$, mass $m$, and the cross-section diameter $d$ all can be taken to be equal to $1$ (with respective units) without any loss of generality, in order to simplify the following expressions. The time-dependent velocity distribution function that is the solution of the BKW problem is given by
\begin{equation}
    f(\mathbf{c},t)= \frac{1}{(\pi C)^{3/2}}\frac{1}{2C}\left(5C - 3 + \frac{2(1-C)c^2}{C}\right)\exp(-c^2/C),\label{eq:bkw-f-an}
\end{equation}
where $C=1-\frac{2}{5} \exp(-t/6)$ is defined for all $t \geq 0$.
Defining a moment of order $2l$ the distribution function as
\begin{equation}
    M^{2l}(t) = \int c^{2l} f(\mathbf{c},t) d\mathbf{c},\label{eq:moment-def}
\end{equation}
it is easily obtained that for the case of the BKW distribution, the analytical expression for moment $M^{2l}$ reads
\begin{equation}
    M^{2l}_{an}(t) = 2C^{l-1}\frac{(2l+2)!}{4^{l+1} (l+1)!}\left(l+C(1-l)\right)\label{eq:bkw-mom-an}.
\end{equation}
Scaling the moments by their corresponding values for a Maxwellian distribution (so that as $t$ tends to $\infty$, the scaled moments all tend to 1), one obtains the following expression for the scaled moment $\hat{M}^{2l}$:
\begin{equation}
    \hat{M}^{2l}(t) = M^{2l}(t) \frac{(l+1)!4^{l+1}}{2(2l+2)!}.
\end{equation}
When performing simulations and comparing their results, several quantities of interest are considered. Due to the stochastic nature of the collision algorithm, noise is present in the simulation, the amount of which is governed by the choice of the $C_{RMS}$ parameter. However, directly comparing results obtained with the discrete velocity method with those given by the analytic solution could conflate the noise present in the simulation with the bias present in the simulation. The bias is due to 1) the discrete nature of the velocity function representation (i.e. grid extent and grid spacing) 2) the remapping procedure 3) the low-rank approximation of the velocity distribution function. Therefore, in order to separate these effects and study them separately, multiple simulations are performed for a given set of initial parameters (grid spacing, $C_{RMS}$, rank of the tensor decomposition). The values of the moments are ensemble-averaged and compared to the analytic solution in order to study the bias in the solution, whereas the standard deviation across the ensemble is computed in order to study the noise in the solution.

In the simulations performed below, the velocity space extent was fixed at $\pm3.5  v_{th}$ in each direction, where $v_{th}=\sqrt{{2kT}/{m}}$ is the thermal speed. Two velocity grids were considered: a ``coarse'' grid with $16^3$ points and a ``fine'' grid with $32^3$ points. The following values of the tensor rank were considered: 4,6,8,10,12 for the coarse grid, and 4,8,12,16,20 for the fine grid. 5 different values of $C_{RMS}$ were used for each simulation: $C_{RMS}=1.5 \cdot 10^{-3}$, $C_{RMS}=5 \cdot 10^{-3} $, $C_{RMS}=10^{-2}$, $C_{RMS}=5 \cdot 10^{-2} $, $C_{RMS}=10^{-1} $. The timestep was taken to be equal to 0.05 of the mean collision time, given by $\tau_c = 1/\left(n \sigma_r v_{th}\right)$, where $n$ is the number density, $\sigma_r=\pi d^2$ is a reference cross-section, and $d$ is the cross-section diameter.
 Thus, 25 sets of simulations were performed using the combined QUIPS/Tensor-Train approach for each velocity grid (5 $C_{RMS}$ values, 5 tensor ranks), and 5 pure sets of QUIPS simulations were performed for each velocity grid (5 $C_{RMS}$ values). In order to obtain mean values and noise levels in the unsteady simulation, ensemble averaging was used, with 250 simulations  performed for each set of parameters. Each simulation was carried out for 200 timesteps (up to a final time of $10\tau_c$).

The simulation bias (for a given set of simulations differing only in the random seed) in a moment of order $2l$ can be defined as follows:
\begin{equation}
    \mathcal{B}_{an}\left(\hat{M}^{2l}\right) =\sqrt{ \frac{1}{N_t}  \sum_{t_i} \left(\overline{\hat{M}^{2l}}(t_i) - \hat{M}^{2l}_{an}(t_i)\right)^2 },
\end{equation}
where $N_t$ is the number of timesteps, $t_i = i\Delta t$ is the time at timestep $i$, and $\overline{\hat{M}^{2l}}$ is the ensemble average of the moments computed with the given set of simulation parameters.

Similarly, the noise in the simulation is defined through the Root-Mean-Square Error:
\begin{equation}
    \mathcal{N}\left(\hat{M}^{2l}\right) =\sqrt{ \frac{1}{N_{ens}}\frac{1}{N_t} \sum_{e=1}^{N_{ens}}\sum_{t_i} \left(\hat{M}^{2l}_{e}(t_i) - \overline{\hat{M}^{2l}}(t_i)\right)^2},
\end{equation}
where $N_{ens}$ is the number of simulation in the ensemble, and the summation over $e$ is the averaging over all the simulations in the ensemble.

\begin{figure}[h!]
    \centering
    \includegraphics[width=.65\textwidth]{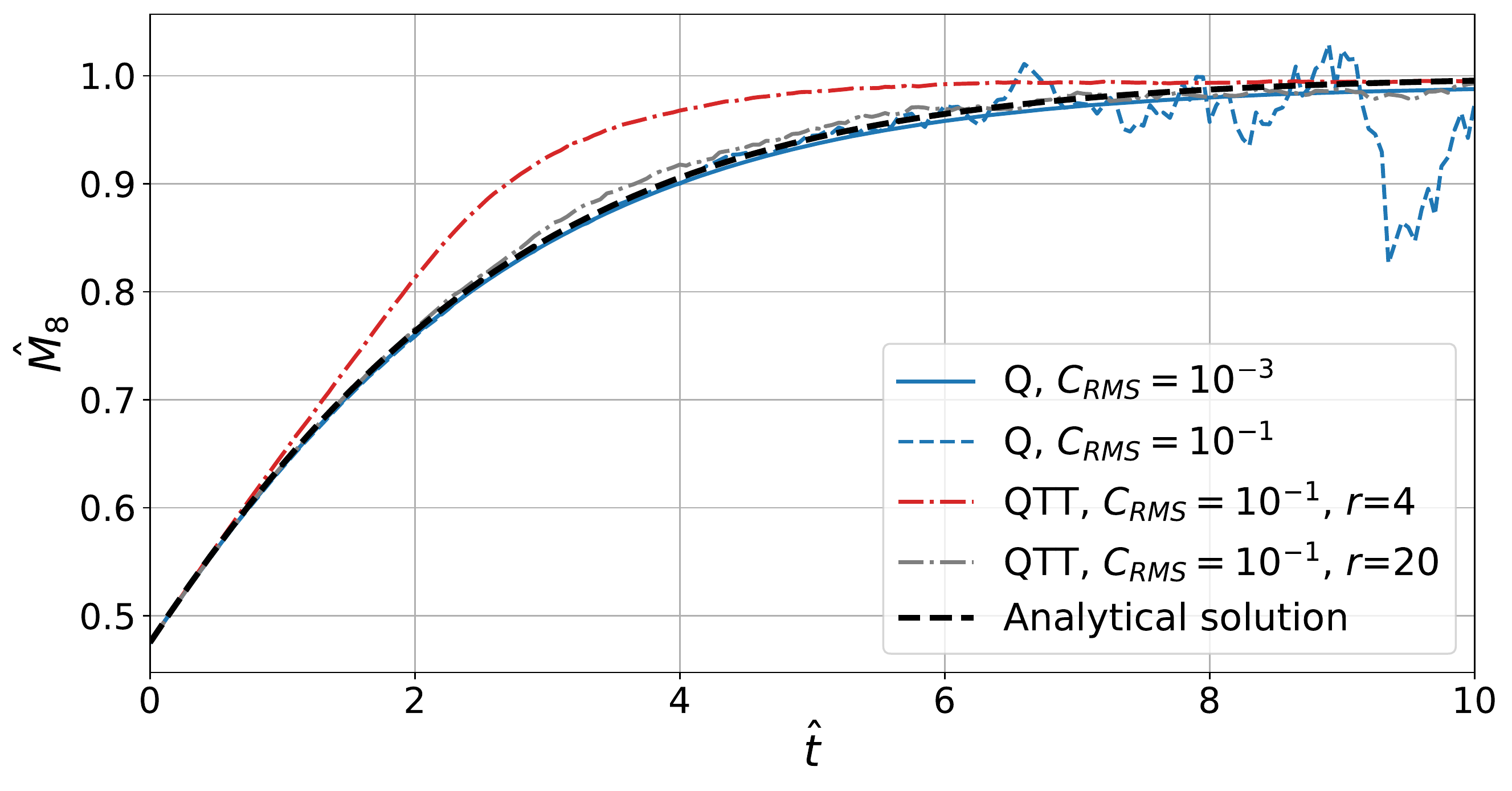}
    \caption{Evolution of the scaled 8$^{th}$ moment for the BKW problem. Pure QUIPS solution denoted by ``Q'', QUIPS/Tensor-Train solution denoted by ``QTT''. 32$^3$ velocity grid.}
    \label{fig:bkw_m8}
\end{figure}

Figure~\ref{fig:bkw_m8} shows the evolution of the scaled 8$^{th}$ moment of the velocity distribution function in time, with the analytic solution shown by the black dashed line, the two pure QUIPS simulations (low-noise with $C_{RMS}=10^{-3}$ and high-noise with $C_{RMS}=10^{-1}$) shown by the solid and dashed blue lines, and two high-noise ($C_{RMS}=10^{-1}$) QUIPS/Tensor-Train solutions with different decomposition ranks shown by the dot-dashed red and grey lines. While the presented figure does not allow for a detailed quantitative analysis and comparison of the different algorithms, it does highlight several important features. One is that at the high value of $C_{RMS}=10^{-1}$, the moments of the compressed distribution function (QTT curves) exhibit significantly less noise compared to the pure QUIPS simulation (Q) curves, even when a relatively large-rank ($r=20$) representation is used, whereas even with ensemble averaging over 250 ensembles, the QUIPS curve exhibits significant noise at the highest $C_{RMS}$ value. For the low-rank ($r=4$) case, the solution is yet smoother, but deviates strongly from the analytic solution and exhibits a faster relaxation to equilibrium. Therefore, the next step is to characterise the performance of the pure QUIPS and QUIPS/Tensor-Train methods in terms of computational cost (as given by computational time and memory use)  and computational error (as given by solution noise and bias).

\begin{figure}[h!]
    \centering
    \includegraphics[width=.92\textwidth]{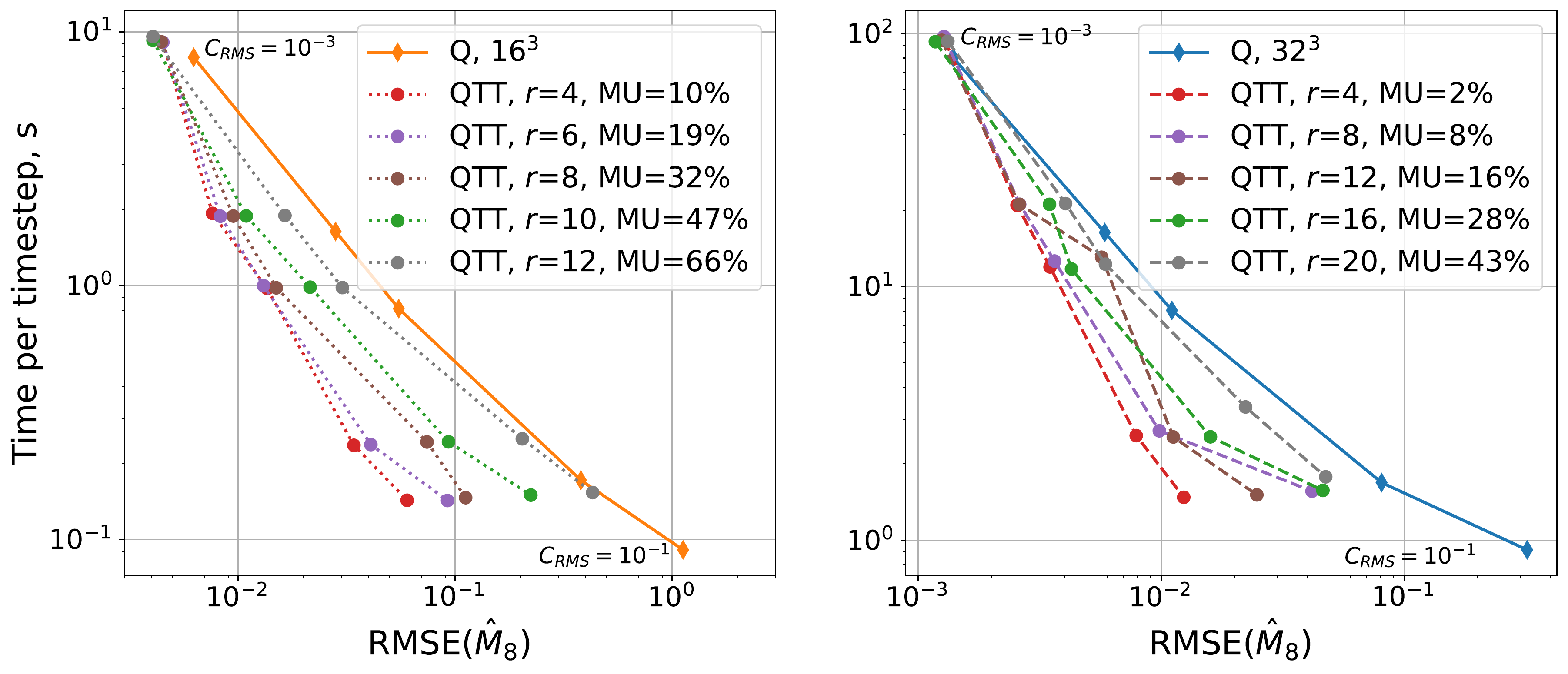}
    \caption{Computational time per timestep plotted against noise in the 8$^{th}$ scaled moment. Pure QUIPS solution denoted by ``Q'', QUIPS/Tensor-Train solution denoted by ``QTT''. $16^{3}$ velocity grid (left), $32^{3}$ velocity grid (right).}
    \label{fig:bkw_rmse_m8}
\end{figure}

Figure~\ref{fig:bkw_rmse_m8} shows the average computational time per timestep plotted against noise in the 8$^{th}$ scaled moment for simulations performed on a coarse $16^{3}$ (left) and fine $32^{3}$ (right) grid. Different QTT curves correspond to different values of the decomposition rank $r$, whereas different points on the curves correspond to different $C_{RMS}$ values, and ``MU`` denotes the machine memory needed to store the Tensor-Train representation of the velocity distribution function as a fraction of the machine memory used by the full (uncompressed) VDF.
A ``good'' stochastic simulation method can be characterized by simultaneously having low computational cost and exhibiting low noise, and a corresponding point on the plot would thus lie in the lower-left corner. As expected, lowering the $C_{RMS}$ values leads to lower noise levels for all the simulations.A reduction in noise is observed when the QUIPS/Tensor-Train algorithm instead of the pure QUIPS approach. The lower the rank of the tensor decomposition, the lower the noise, as the lower-rank representation effectively smooths out the stochastic fluctuations in the distribution function tails. At lower $C_{RMS}$ values the effect is diminished, as the noise levels become sufficiently low so as to not hinder the effectiveness of a lower-rank representation of the distribution function, and thus little artificial smoothing is introduced by the tensor decomposition. At low $C_{RMS}$ values the computational costs are also virtually identical, as the additional overhead introduced by the tensor decomposition and reconstruction of the full VDF from the compressed representation is negligible compared to the overall cost of the collision computation. At high $C_{RMS}$ values use the QUIPS/Tensor-Train algorithm is more expensive (compared to a pure QUIPS simulation performed with the same $C_{RMS}$ value), however, for a given level of computational cost, achieving a similar noise level would require a lower $C_{RMS}$ value in the pure QUIPS approach and a higher cost.

\begin{figure}[h!]
    \centering
    
    \includegraphics[width=.92\textwidth]{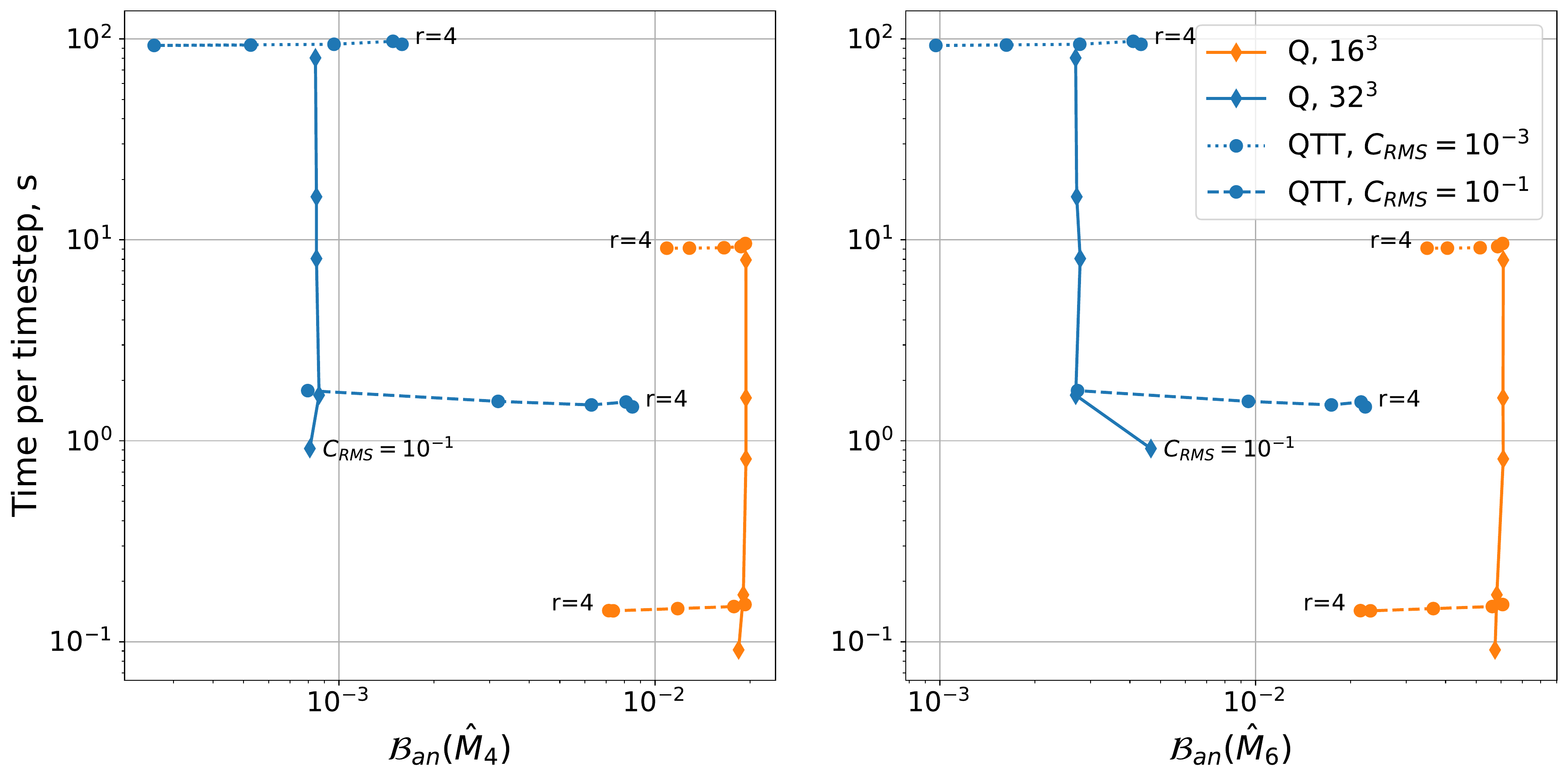}
    \caption{Computational time per timestep plotted against bias in the 4$^{th}$ (left) and 6$^{th}$ (right) scaled moments.}
    \label{fig:bkw_bias}
\end{figure}

\begin{figure}[h!]
    \centering
    
    \includegraphics[width=.65\textwidth]{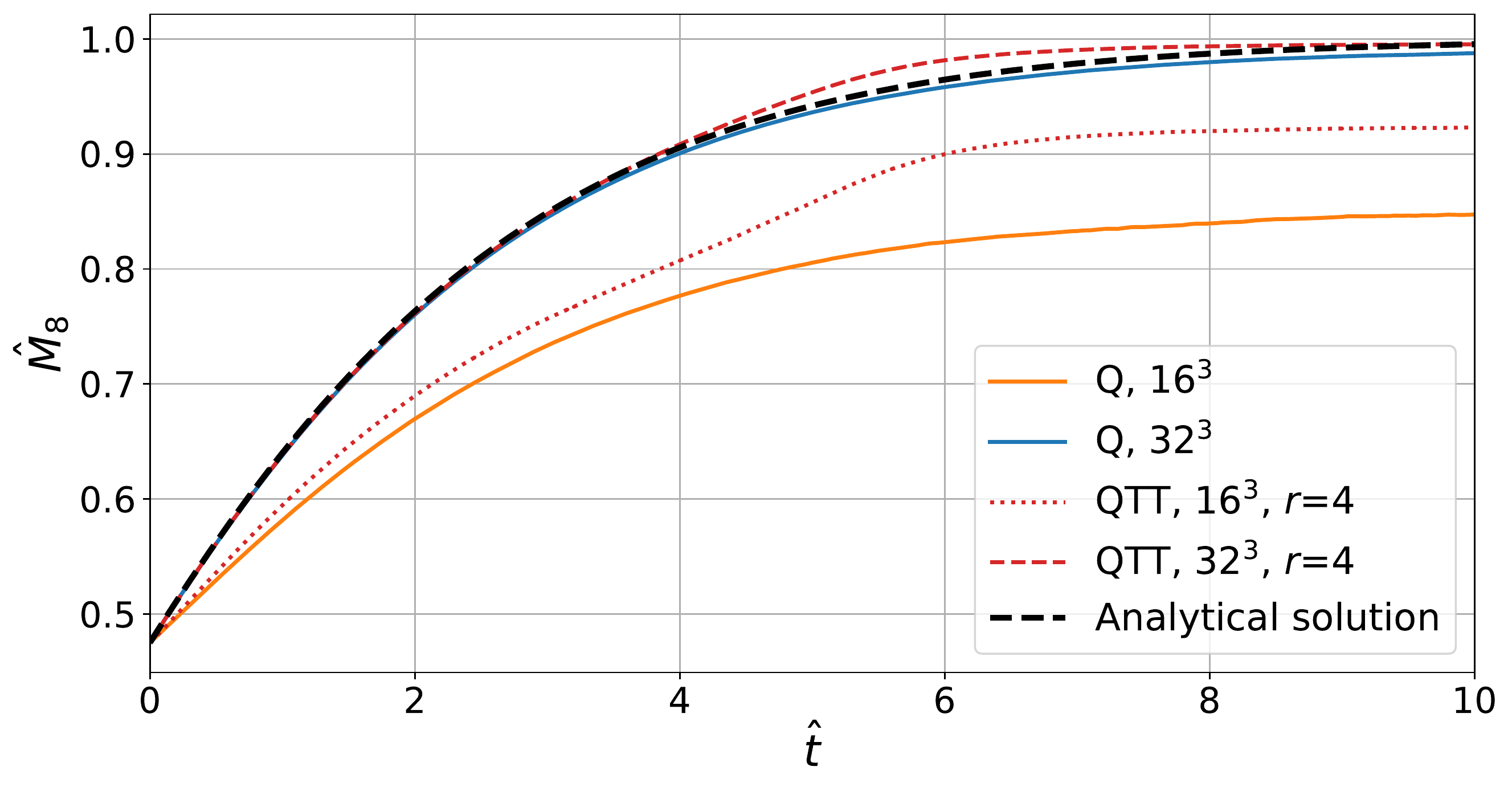}
    \caption{Evolution of the scaled 8$^{th}$ moment for the BKW problem. Fixed $C_{RMS}$ value of $10^{-3}$.}
    \label{fig:bkw_m8_grids}
\end{figure}

Next, we are interested in the bias introduced in the simulations. Two lower-order moments (4$^{th}$ and 6$^{th}$) are considered, as the 8$^{th}$ order moment is noisy at high $C_{RMS}$ values even with ensemble averaging (as seen on Fig.~\ref{fig:bkw_m8}), and thus is not well-suited for analysis of bias compared to the smooth analytic solution. Figure~\ref{fig:bkw_bias} shows the bias in the 4$^{th}$ (left) and 6$^{th}$ (right) scaled moments. The different points on the pure QUIPS curves correspond to different $C_{RMS}$ values, whereas different points on the QUIPS/Tensor-Train curves correspond to different values of the decomposition rank $r$, and the QUIPS/Tensor-Train results are plotted for the lowest ($C_{CRMS}=10^{-3}$) and highest ($C_{RMS}=10^{-1}$) values of the $C_{RMS}$ parameter. For the pure QUIPS simulations (``Q'' curves) it can be observed that use of a finer velocity space grid leads to a significant reduction in the bias, whereas the value of the $C_{RMS}$ parameter has very little influence on the bias in the solution. For the QUIPS/Tensor-Train results, the impact of varying the decomposition rank $r$ varies on the velocity grid used --- for the coarse velocity grid, the lower the rank $r$, the lower the bias, whereas for the fine velocity grid, the lower the rank, the higher the bias. To explain this seemingly contradictory behaviour, the time evolution of the 8$^{th}$ scaled moment for both the coarse and fine velocity grids is plotted on Fig.~\ref{fig:bkw_m8_grids}. It can be seen that use of a coarse decomposition $r=4$ introduces bias compared to the pure QUIPS solution on the same velocity grid, and that this bias is especially strong for the case of the coarse velocity grid. However, due to the bias leading to higher computed values of the distribution function moment, it causes a reduction in the bias computed with reference to the analytic solution. Thus, care has to be taken when analyzing such metrics, as the role of the bias introduced by the velocity space discretization needs to be accounted for as well. 
The amount of bias introduced is also affected by the level of noise in the simulation (as governed by the $C_{RMS}$ value), as higher levels of noise lead to more error in the VDF decomposition, and consequently, a higher bias in the overall solution.

\begin{figure}[h!]
    \centering
    
    \includegraphics[width=.92\textwidth]{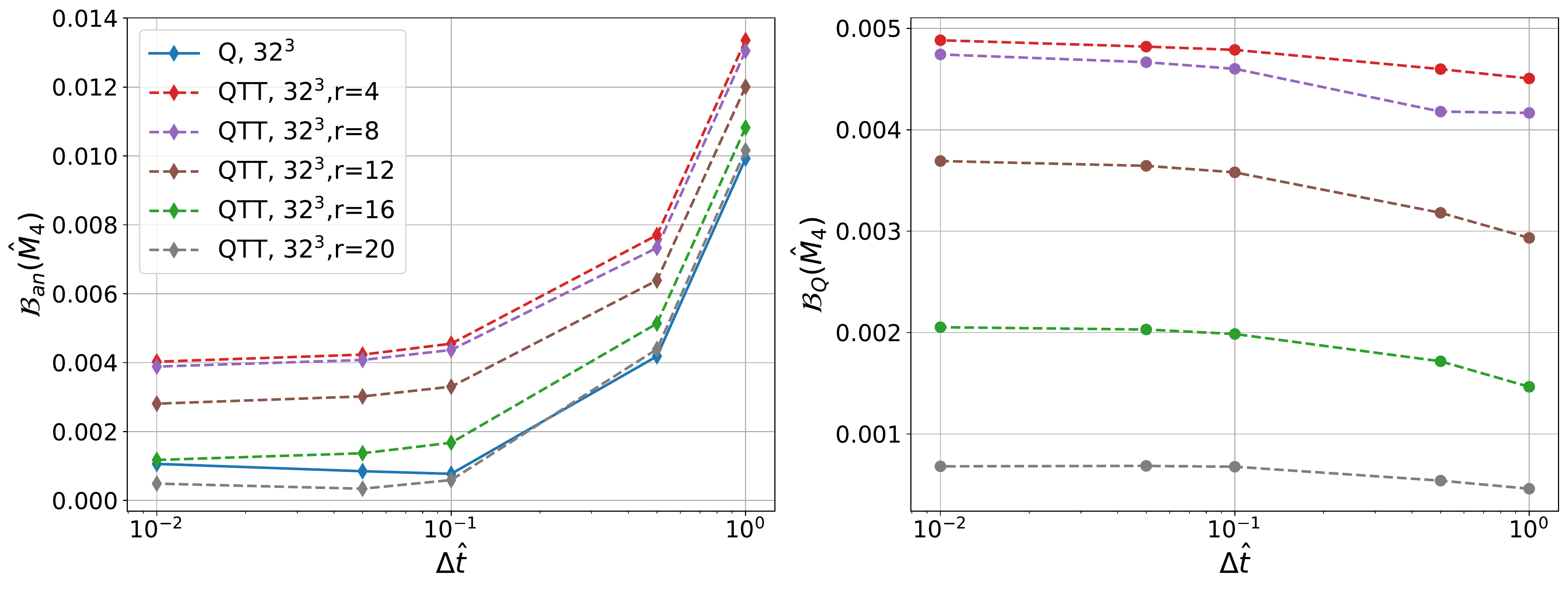}
    \caption{Bias as a function of timestep size with respect to analytic BKW solution (left) and pure QUIPS solution (right).}
    \label{fig:bkw_biasvdt}
\end{figure}

Finally, we are also interested in the choice of timestep on the bias and noise in the simulations. On one hand, a larger timestep leads to a larger error in the estimation of the collision integral. On the other hand, use of a smaller timestep in conjunction with the tensor decomposition method will lead to more frequent compression and reconstruction of the distribution function, and thus can be a cause of additional error, although the lower number of collisions performed during each timestep reduces the amount of noise introduced at each timestep. In order to better quantify the impact of varying the simulation timestep, an additional set of simulations was carried on the ``fine'' $32^3$ velocity grids with a $C_{RMS}$ value of $10^{-2}$, using timesteps of $0.01 \tau_c$, $0.05 \tau_c$, $0.1 \tau_c$, $0.5 \tau_c$, and $1.0 \tau_c$, running all simulations up to a final time of $t=10 \tau_c$.

Figure~\ref{fig:bkw_biasvdt} presents the bias in the $4^{th}$ scaled moment as a function of the timestep used, with the bias computed with respect to the analytic solution (left) and to the ensemble-averaged pure QUIPS solution (right).  From the left subplot it can be seen that the behaviour of both the pure QUIPS and QUIPS/Tensor-Train algorithms is very similar: use of timesteps smaller than $0.1$ mean collision time has little impact on the bias, whereas using larger timesteps leads to a strong correlation between the bias and timestep size due to error in evaluation of the collision integral. To analyze the impact of choice of timestep on the tensor decomposition-based algorithm, the bias with respect to the pure QUIPS solution is plotted (Fig.~\ref{fig:bkw_biasvdt}, right subplot). The bias can be seen to be very weakly dependent on the choice of timestep as long as the timestep is lower than $0.1$ of the mean collision time; for larger timesteps some reduction in bias compared to the reference pure QUIPS solution can be observed, although it should be noted that the QUIPS solution itself suffers from a large error at such large timesteps.

Thus, concluding the analysis of the presented BKW simulations, it can be stated that use of the Tensor-Train decomposition has a smoothing effect on the distribution function, leading to a reduction of noise in the simulations, especially at higher initial noise levels; moreover, the computational cost required to achieve similar noise levels with the pure QUIPS approach would have been higher than when using the QUIPS/Tensor-Train algorithm. However, the tensor decomposition introduces bias into the simulation, the magnitude of which is dependent not only the rank of the decomposition used, but also on the level of noise in the simulation (with higher levels of noise leading to higher bias levels). The amount of bias introduced in the simulation by the tensor decomposition was found to be weakly dependent on the timestep as long as reasonably small timesteps (0.1 of the mean collision time or lower) were used, which is not qualitatively different from the collision timestep restriction on the pure QUIPS simulation approach.

\subsection{Couette flow}
Next, a 1-dimensional Couette flow of argon in a channel with a width of 0.25 mm is considered. The wall temperatures $T_w$ were taken to be 300~K, and the y-velocities of the left and right wall were taken to be $-500$  m/s and $500$ m/s, respectively. The gas number density was taken to be 10$^{23}$ m$^{-3}$, corresponding to a Knudsen number of approximately 0.05. 150 cells were used to discretize the domain, and a 24$^3$ velocity grid was used, with an extent of $\left[-4.5v_{ref},4.5v_{ref}\right]$, where $v_{ref} = \sqrt{2kT_{w} / m}$, and $m$ is the argon atom mass. The VHS (Variable Hard Sphere~\cite{bird1994molecular}) cross-section model was used for collisions, with a diameter of 4.11~\AA~and a VHS cross-section velocity exponent $\omega=0.81$. 
The following values of $C_{RMS}$ were considered: 10$^{-2}$, 2.5$\cdot 10^{-2}$, and 7.5$\cdot 10^{-2}$. The ranks of the reduced-order representations considered were taken to be 4, 8, 12, and 16. The simulations were carried out for 50000 timesteps, with time-averaging performed after 20000 timesteps had elapsed and the flow had reached a steady state.

\begin{figure}[h!]
    \centering
    
    \includegraphics[width=.92\textwidth]{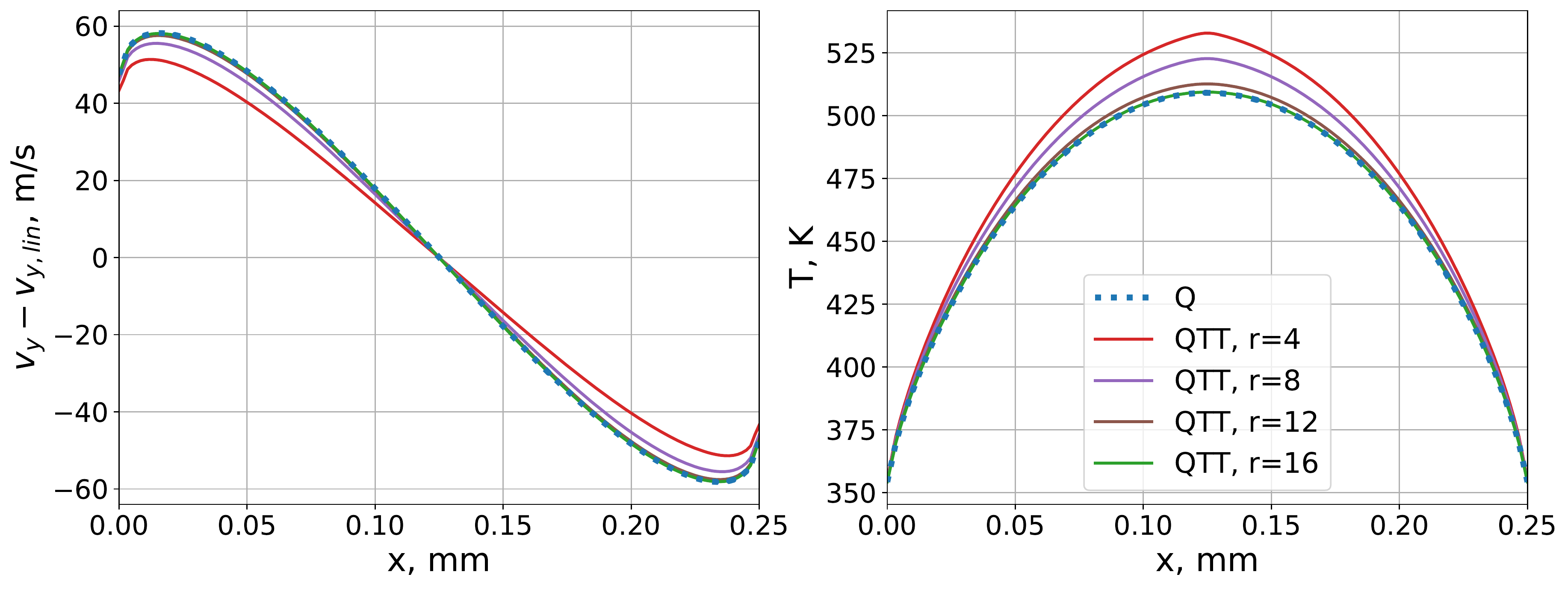}
    \caption{Time-averaged profiles of deviation of velocity from the linear profile (left) and of the temperature (right). $C_{RMS}$ value of 10$^{-2}$.}
    \label{fig:couette_vT_1e2}
\end{figure}

Figure~\ref{fig:couette_vT_1e2} shows the time-averaged profiles of the deviation of the computed y-velocity from the linear velocity profile given by $v_{y,lin}(x)=-500 + 1000 x / 0.25$ (where $x$ is in mm) (left) and the flow temperature (right). It can be seen that the tensor decomposition introduces bias in both the velocity and temperature profiles, with lower-order decomposition ranks leading to higher bias (decrease in the velocity slip magnitude and increase in the temperature), consistent with the behaviour seen in the BKW relaxation case. As in the BKW case, it also expected that at higher $C_{RMS}$ values, the bias due to the tensor decomposition will be higher due to the noise negatively affecting the lower-rank VDF representation.

\begin{figure}[h!]
    \centering
    
    \includegraphics[width=.92\textwidth]{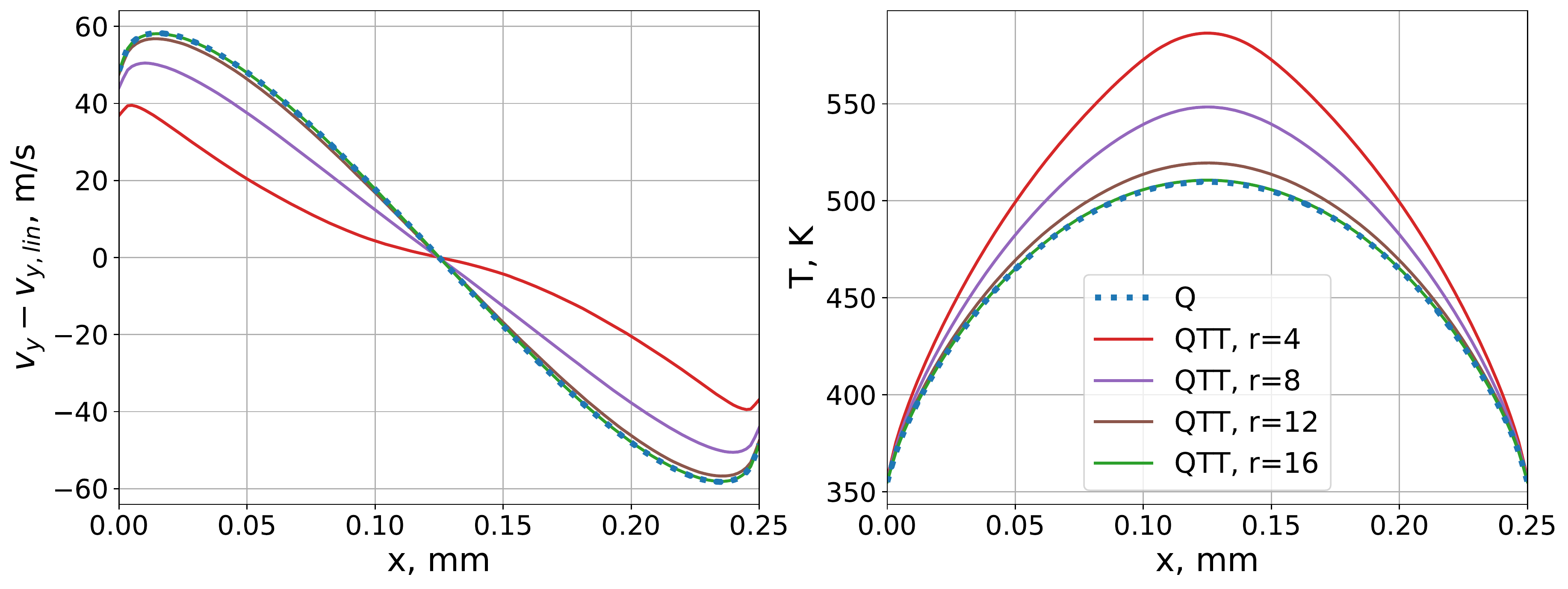}
    \caption{Time-averaged profiles of deviation of velocity from the linear profile (left) and of the temperature (right). $C_{RMS}$ value of 7.5$\cdot 10^{-2}$.}
    \label{fig:couette_vT_75e2}
\end{figure}

Figure~\ref{fig:couette_vT_75e2} shows the same quantities as Fig.~\ref{fig:couette_vT_1e2}, but computed with a $C_{RMS}$ value of $7.5 \cdot 10^{-2}$. Thus, as predicted, a much more significant increase in bias with decreasing $r$ is observed. However, the $r=16$ and $r=12$ rank decompositions still provide a good agreement with the pure QUIPS results, and yet require approximately 50\% and 70\% less memory to store the compressed VDF, correspondingly.

\begin{figure}[h!]
    \centering
    
    \includegraphics[width=.92\textwidth]{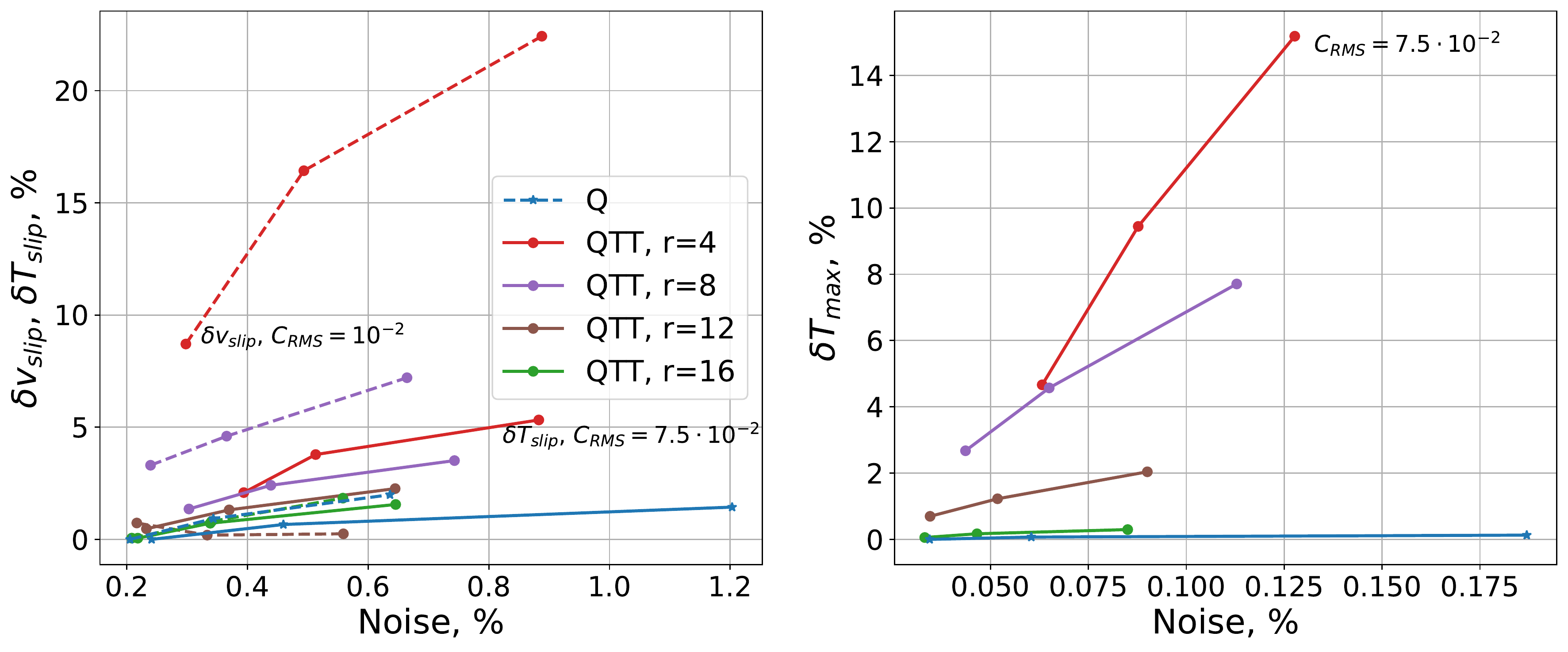}
    \caption{Bias in slip velocity and temperature (left) and maximum temperature (right) plotted against noise in the same quantities. Different points correspond to different $C_{RMS}$ values.}
    \label{fig:couette_biaserror}
\end{figure}

Figure~\ref{fig:couette_biaserror} shows the noise and bias in the different quantities: the slip velocity (left subplot, dashed lines) and temperature (left subplot, solid lines) and maximum value of the temperature (right subplot) for different values of $r$ and $C_{RMS}$. To compute the bias, the time-averaged pure QUIPS solution computed with $C_{RMS}=10^{-2}$ was used as a reference solution. Different points on the curves correspond to different $C_{RMS}$ values. As expected, varying $C_{RMS}$ has little impact on the bias in the pure QUIPS results, whereas its effect becomes more pronounced as the tensor decomposition rank $r$ is decreased.

\begin{figure}[h!]
    \centering
    
    \includegraphics[width=.92\textwidth]{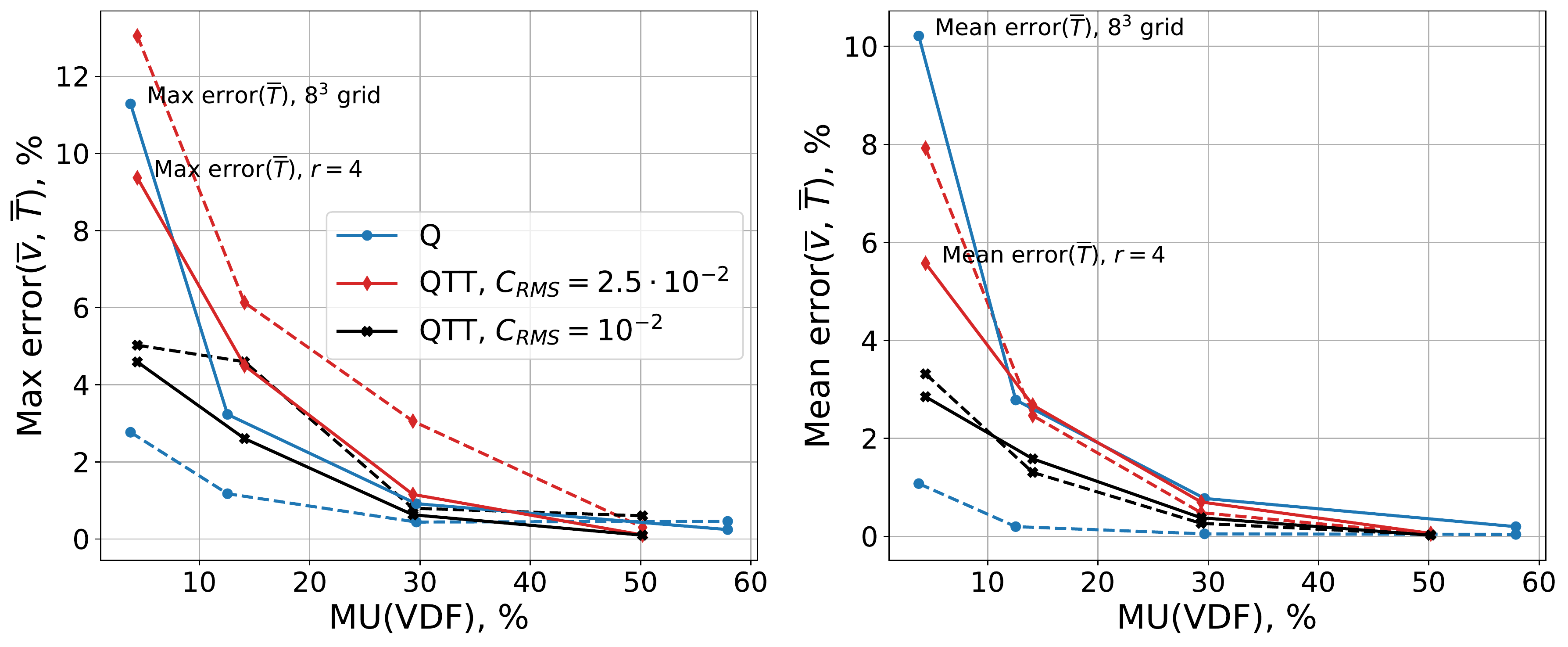}
    \caption{Maximum relative deviation (left) and root-mean-square relative deviation (right) of velocity (dashed lines) and temperature (solid lines) plotted against memory use with respect to the full 24$^3$ QUIPS solution.}
    \label{fig:couette_errmemuse}
\end{figure}

Finally, it is of interest to investigate which method performs better: pure QUIPS on a coarse velocity grid or QUIPS/Tensor-Train on a finer grid with a low decomposition rank?
To this extent, additional pure QUIPS simulations were conducted, with velocity grid resolutions chosen so as to be close in terms of memory use to the compressed Tensor-Train representations. The grid sizes chosen were 8$^3$, 12$^3$, 16$^3$, and 24$^3$.

Of course, using a coarser velocity grid discretization entails not only a decrease in the memory use, but also in the computational cost due to both the convection and the collisions (as the velocity grid $\Delta v$ spacing enters the equation for the number of collisions ~(\ref{eq:ncoll})). For a more consistent comparison, the $C_{RMS}$ values were adjusted so as to produce approximately the same number of collisions at the start of the simulation as the pure QUIPS solution on a $24^3$ grid with a $C_{RMS}$ value of $2.5 \cdot 10^{-2}$.
Figure~\ref{fig:couette_errmemuse} shows the maximum (left) and mean (right) relative deviations in the velocity (dashed line) and temperature (solid line) with respect to the QUIPS solution on a $24^3$ velocity grid (which was considered as a reference solution) for different grid sizes and decomposition ranks, represented on the X-axis by the memory use (in percent) of the lower-resolution VDF compared to storing the full $24^3$ VDF. It can be seen that at lower $C_{RMS}$ values (black QTT curves) the TensorTrain-based representation performs better than pure QUIPS on a coarse velocity grid, especially in the low memory-use cases, as the error in the temperature due to the grid discretization becomes quite significant. However, for the error in the velocity, the pure QUIPS approach retains the lowest error of all the results presented.

Thus, it can be concluded that for the supersonic Couette flow test-case, the QUIPS/Tensor-Train approach can achieve errors of less than 2\% in the velocities and temperatures, whilst reducing the memory cost by 50\% ($r=16$) to 30\% ($r=12$), provided that low $C_{RMS}$ values are used; however, the approach does not out-perform the pure QUIPS approach unless the latter uses very low-resolution velocity grids.

\subsection{Fourier flow}
Finally, a 1-dimensional Fourier flow of argon in a channel with a width of 0.25 mm is considered, with a left wall temperature of 300~K and a right wall temperature of 600~K. All numerical parameters were the same as the Couette test case, except for the velocity grid extent, which was taken to be $\left[-5v_{ref},5v_{ref}\right]$, where $v_{ref} = \sqrt{2kT_{w,l} / m}$, where $T_{w,l}$ is the temperature of the left wall.

\begin{figure}[h!]
    \centering
    
    \includegraphics[width=.92\textwidth]{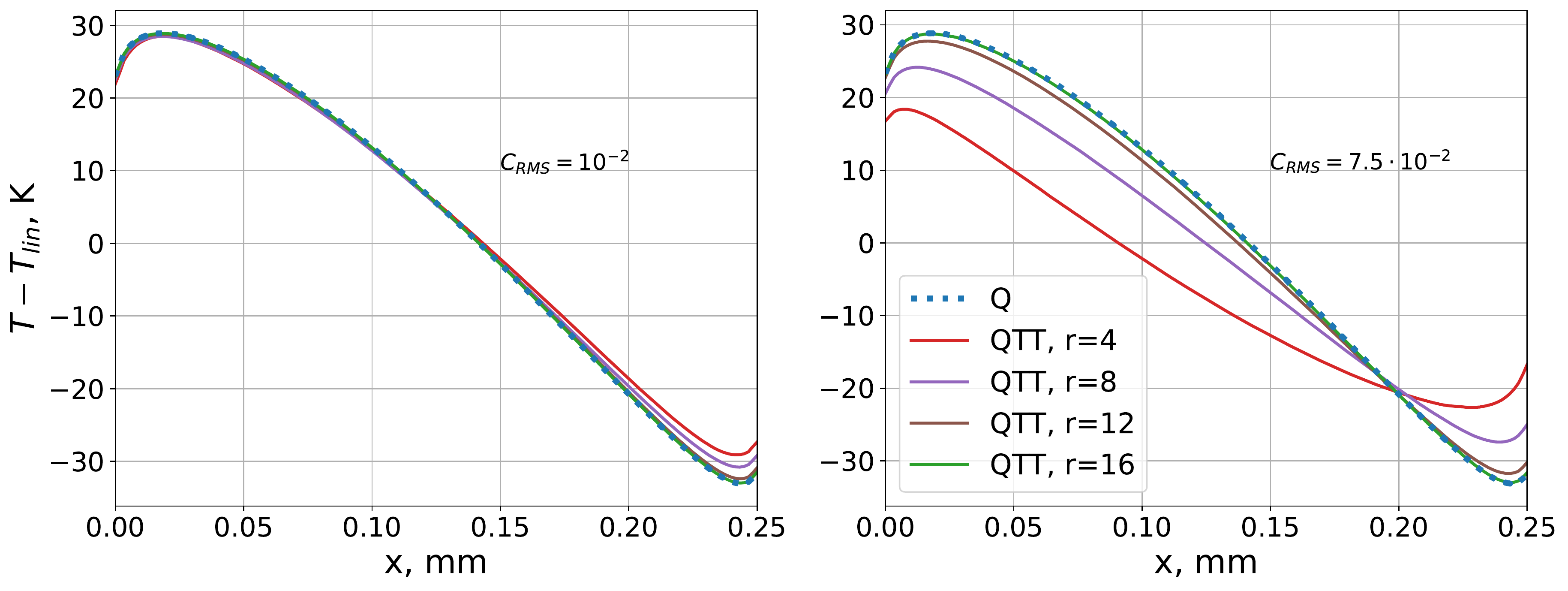}
    \caption{Time-averaged profiles of deviation of temperature from the linear profile. $C_{RMS}$ values of 10$^{-2}$ (left) and $7.5 \cdot 10^{-2}$ (right).}
    \label{fig:fourier_vT_1e2}
\end{figure}

Figure~\ref{fig:fourier_vT_1e2} shows the time-averaged profiles of the deviation of the flow temperature from that given by a linear profile $T_{lin}(x)=300 + 300 x / 0.25$, where $x$ is in mm. As in previous situations, at a low $C_{RMS}$ value (left subplot), all solutions provide similar results, but as the noise level is increased, lower-rank decompositions exhibit a significant error (right subplot). For the case of the Fourier flow, the temperature (and temperature slip at the right wall) and the orthogonal component of the heat flux at the right wall are considered as quantities of interest. The quantities at the colder (left) wall exhibit the same tendencies quantitatively and qualitatively as those at the right wall, and are thus not presented.

\begin{figure}[h!]
    \centering
    
    \includegraphics[width=.92\textwidth]{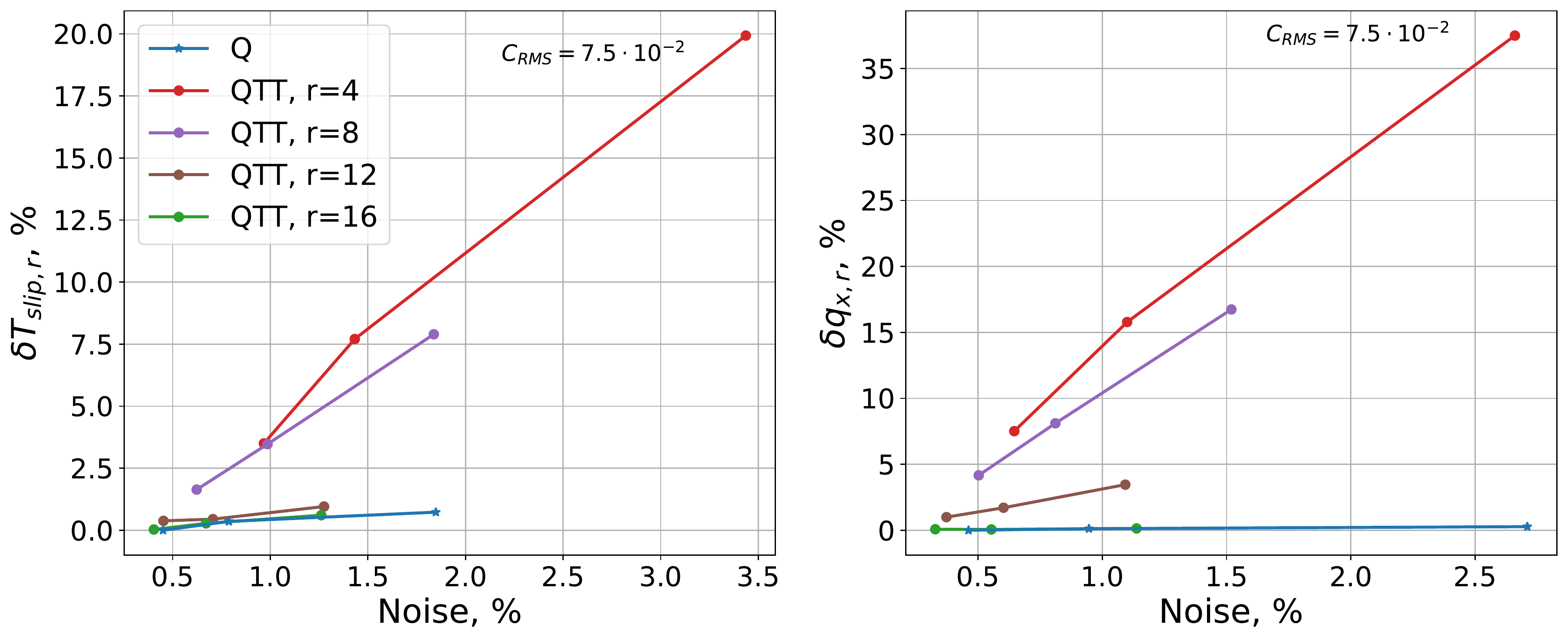}
    \caption{Bias in slip temperature (left) and heat flux (right) plotted against noise in the same quantities. Different points correspond to different $C_{RMS}$ values.}
    \label{fig:fourier_biaserror}
\end{figure}

Figure~\ref{fig:fourier_biaserror} shows the bias (computed with respect to the lowest-noise pure QUIPS solution) and noise in the slip temperature at the right wall (left) and heat flux (right). As in the case of the Couette flow, the bias is strongly dependent on both the noise level and decomposition rank, with a stronger $C_{RMS}$ depends in case of stronger compression. As could be expected, the bias is higher for the higher-order moment (heat flux), however, the rank 12 and 16 decompositions achieve results that are very close to the pure QUIPS approach.

\begin{figure}[h!]
    \centering
    
    \includegraphics[width=.92\textwidth]{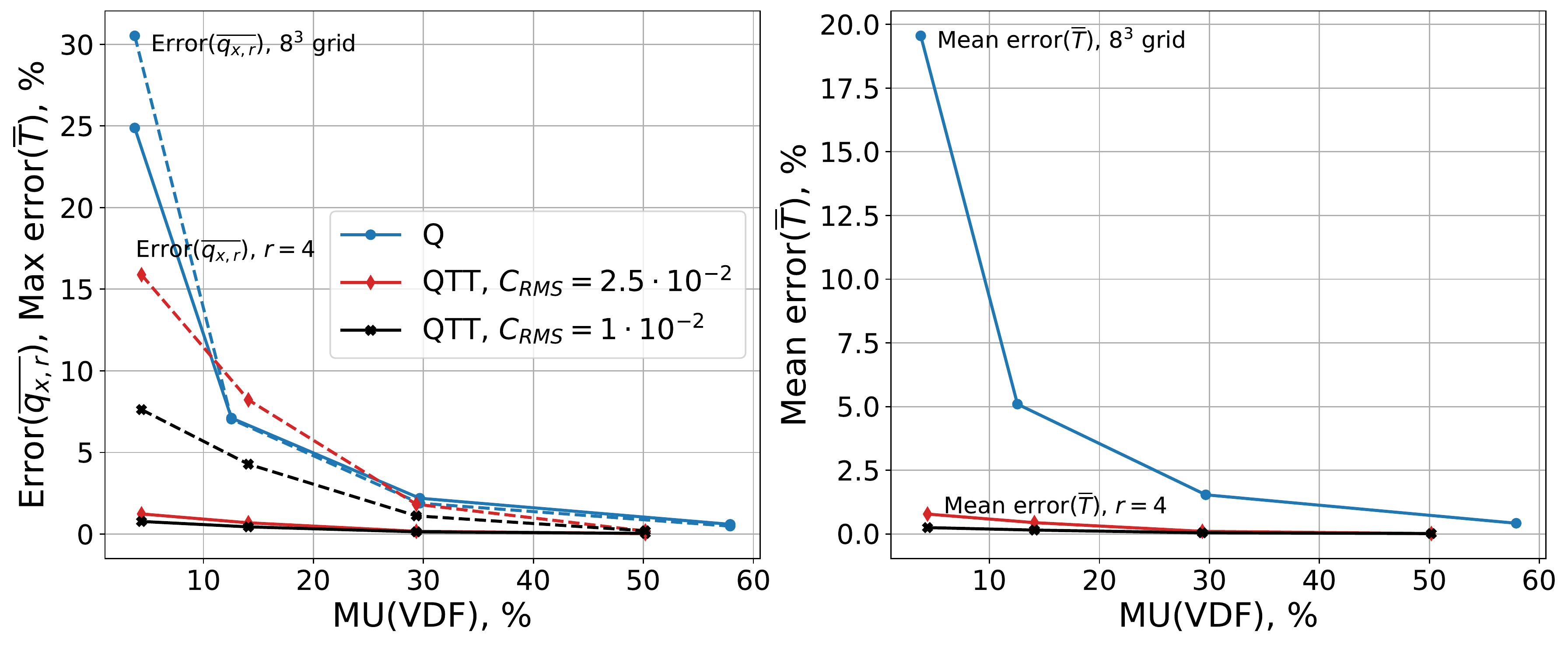}
    \caption{Maximum relative deviation of temperature (solid lines) and heat flux (dashed lines) (left) and root-mean-square relative deviation of temperature (right)  plotted against memory use with respect to the full 24$^3$ QUIPS solution.}
    \label{fig:fourier_errmemuse}
\end{figure}

Finally, as for the Couette test case, an analysis is carried out of how the low-rank tensor decomposition-based solutions compare to the pure QUIPS algorithm on a coarse velocity grid. Fig.\ref{fig:fourier_errmemuse} shows the error in the heat flux at the right wall and the maximum error in the temperature (left; error in the heat flux shown with dashed lines, error in the temperature shown with solid lines) and mean error in the temperature (right), computed with respect to the pure QUIPS solution on a 24$^3$ grid. For the temperature, the QUIPS/Tensor-train approach provides a much lower error than the pure QUIPS approach for both noise levels considered; and for the heat flux, at lower $C_{RMS}$ values, the tensor decomposition-based algorithm also achieves a smaller error in the computed heat flux. Compared to the Couette flow test case, here, at the lowest machine memory use values, the tensor decomposition based algorithm provides a marked improvement of the pure QUIPS approach.

\section{Conclusions}
\label{sec:conclusions}
An energy-conserving modification of the Tensor-Train tensor decomposition algorithm for rarefied gas dynamics applications has been developed and applied to a stochastic discrete velocity-based simulation method, the ``Quasi-Particle Simulation method'' (QUIPS). An interleaved collision/convection algorithm has also been proposed, that allows for simple application of a spatially second-order convection scheme.
The errors (in terms of both solution noise and bias) introduced by the use of the tensor decomposition in conjunction with the stochastic collision evaluations have been analyzed for the 0-D BKW relaxation, and 1-D Couette and Fourier flow problems.
It has been shown that noise in the VDF introduces additional bias when the Tensor-Train reduced-order representation is used, especially when lower decomposition ranks are utilized. However, unless very low-rank decompositions and/or high noise levels are used, the results obtained with the hybrid QUIPS/Tensor-Train approach can be quite accurate in terms of noise and bias, whilst simultaneously reducing the memory cost compared to a pure QUIPS approach. The conservative Tensor-Train tensor decomposition/collision algorithm has also been shown to introduce minimal additional computational cost, unless very high noise levels are present in the simulation. The proposed algorithm was also shown to be weakly influenced by the choice of timestep.

In the case of the Couette and Fourier flows, the use of decompositions with relatively high ranks provides results very similar to those achieved by the pure QUIPS approach, whilst allowing for 50\% to 70\% reduction in the storage requirements for the VDF in a single physical cell. For stronger compression (lower decomposition ranks), the QUIPS/Tensor-Train approach provides less error than the pure QUIPS approach does for the same memory cost. As such, the energy-conserving QUIPS/Tensor-Train algorithm has been shown to be a viable approach to the reduction of memory costs in rarefied flow simulations, even in the presence of stochastic noise in the solution.

Possible future extensions to the method include 1) extension to multi-species flows 2) extension to molecular flows with ro-vibrational distributions~\cite{clarke2018low}, where the potential memory savings would be even larger 3) extension to distributions defined on spherically cut-off velocity grids~\cite{poondla2022modeling} (or in general, extension to non-Cartesian velocity space coordinate systems) 4) inclusion of conservative projection methods and 5) combination with machine-learning based methods for faster evaluation of the collision integral.



\section*{Acknowledgments}
Georgii Oblapenko acknowledges the funding provided by the Alexander von Humboldt foundation for his stay as a guest researcher at the German Aerospace Center (DLR).


\begin{thebibliography}{10}

\bibitem{bird1994molecular}
G.~A. Bird, ``Molecular gas dynamics and the direct simulation of gas flows,''
  {\em Molecular gas dynamics and the direct simulation of gas flows}, 1994.

\bibitem{platkowski1988discrete}
T.~Platkowski and R.~Illner, ``Discrete velocity models of the {Boltzmann}
  equation: a survey on the mathematical aspects of the theory,'' {\em SIAM
  review}, vol.~30, no.~2, pp.~213--255, 1988.

\bibitem{filbet2006solving}
F.~Filbet, C.~Mouhot, and L.~Pareschi, ``Solving the {Boltzmann} equation in
  {N} log2 {N},'' {\em SIAM Journal on Scientific Computing}, vol.~28, no.~3,
  pp.~1029--1053, 2006.

\bibitem{gamba2017fast}
I.~M. Gamba, J.~R. Haack, C.~D. Hauck, and J.~Hu, ``A fast spectral method for
  the {Boltzmann} collision operator with general collision kernels,'' {\em
  SIAM Journal on Scientific Computing}, vol.~39, no.~4, pp.~B658--B674, 2017.

\bibitem{hara2012one}
K.~Hara, I.~D. Boyd, and V.~I. Kolobov, ``One-dimensional hybrid-direct kinetic
  simulation of the discharge plasma in a {Hall} thruster,'' {\em Physics of
  Plasmas}, vol.~19, no.~11, p.~113508, 2012.

\bibitem{chan2022enabling}
W.~H.~R. Chan and I.~D. Boyd, ``Enabling direct kinetic simulation of dense
  plasma plume expansion for laser ablation plasma thrusters,'' {\em Journal of
  Electric Propulsion}, vol.~1, no.~1, p.~26, 2022.

\bibitem{chan2023grid}
W.~H.~R. Chan and I.~D. Boyd, ``Grid-point requirements for direct kinetic
  simulation of weakly collisional plasma plume expansion,'' {\em Journal of
  Computational Physics}, vol.~475, p.~111861, 2023.

\bibitem{mieussens2000discrete}
L.~Mieussens, ``Discrete-velocity models and numerical schemes for the
  {Boltzmann-BGK} equation in plane and axisymmetric geometries,'' {\em Journal
  of Computational Physics}, vol.~162, no.~2, pp.~429--466, 2000.

\bibitem{dimarco2014numerical}
G.~Dimarco and L.~Pareschi, ``Numerical methods for kinetic equations,'' {\em
  Acta Numerica}, vol.~23, pp.~369--520, 2014.

\bibitem{bgk}
P.~L. Bhatnagar, E.~P. Gross, and M.~Krook, ``A model for collision processes
  in gases. \protect{I}. small amplitude processes in charged and neutral
  one-component systems,'' {\em Physical Review}, vol.~94, no.~3, p.~511, 1954.

\bibitem{holway1966new}
L.~H. Holway~Jr, ``New statistical models for kinetic theory: methods of
  construction,'' {\em The physics of fluids}, vol.~9, no.~9, pp.~1658--1673,
  1966.

\bibitem{shakhov1968generalization}
E.~Shakhov, ``Generalization of the {Krook} kinetic relaxation equation,'' {\em
  Fluid dynamics}, vol.~3, no.~5, pp.~95--96, 1968.

\bibitem{bisi2021kinetic}
M.~Bisi and R.~Travaglini, ``A kinetic {BGK} relaxation model for a reacting
  mixture of polyatomic gases,'' in {\em Recent Advances in Kinetic Equations
  and Applications}, pp.~69--92, Springer, 2021.

\bibitem{mathiaud2022bgk}
J.~Mathiaud, L.~Mieussens, and M.~Pfeiffer, ``An es-bgk model for diatomic
  gases with correct relaxation rates for internal energies,'' {\em European
  Journal of Mechanics-B/Fluids}, vol.~96, pp.~65--77, 2022.

\bibitem{haack2023numerical}
J.~Haack, C.~D. Hauck, C.~F. Klingenberg, M.~Pirner, and S.~Warnecke,
  ``Numerical schemes for a multi-species {BGK} model with velocity-dependent
  collision frequency,'' {\em Journal of Computational Physics}, vol.~473,
  p.~111729, 2023.

\bibitem{xu2010unified}
K.~Xu and J.-C. Huang, ``A unified gas-kinetic scheme for continuum and
  rarefied flows,'' {\em Journal of Computational Physics}, vol.~229, no.~20,
  pp.~7747--7764, 2010.

\bibitem{pfeiffer2022exponential}
M.~Pfeiffer, F.~Garmirian, and M.~Gorji, ``Exponential {Bhatnagar-Gross-Krook}
  integrator for multiscale particle-based kinetic simulations,'' {\em Physical
  Review E}, vol.~106, no.~2, p.~025303, 2022.

\bibitem{tcheremissine1998conservative}
F.~Tcheremissine, ``Conservative evaluation of {Boltzmann} collision integral
  in discrete ordinates approximation,'' {\em Computers \& Mathematics with
  Applications}, vol.~35, no.~1-2, pp.~215--221, 1998.

\bibitem{morris2011monte}
A.~Morris, P.~L. Varghese, and D.~B. Goldstein, ``Monte carlo solution of the
  {Boltzmann} equation via a discrete velocity model,'' {\em Journal of
  Computational Physics}, vol.~230, no.~4, pp.~1265--1280, 2011.

\bibitem{tcheremissine2012method}
F.~Tcheremissine, ``Method for solving the {Boltzmann} kinetic equation for
  polyatomic gases,'' {\em Computational Mathematics and Mathematical Physics},
  vol.~52, pp.~252--268, 2012.

\bibitem{clarke2018low}
P.~Clarke, P.~Varghese, and D.~Goldstein, ``A low noise discrete velocity
  method for the {Boltzmann} equation with quantized rotational and vibrational
  energy,'' {\em Journal of Computational Physics}, vol.~352, pp.~326--340,
  2018.

\bibitem{poondla2022modeling}
Y.~Poondla, D.~Goldstein, P.~Varghese, P.~Clarke, and C.~Moore, ``Modeling
  rarefied gas chemistry with {QuiPS}, a novel quasi-particle method,'' {\em
  Theoretical and Computational Fluid Dynamics}, pp.~1--36, 2022.

\bibitem{oblapenko2021velocity}
G.~Oblapenko, D.~Goldstein, P.~Varghese, and C.~Moore, ``Velocity-space
  hybridization of {Direct Simulation Monte Carlo} and a {Quasi-Particle
  Boltzmann} solver,'' {\em Journal of Thermophysics and Heat Transfer},
  vol.~35, no.~4, pp.~788--799, 2021.

\bibitem{oblapenko2021modeling}
G.~Oblapenko, D.~B. Goldstein, P.~Varghese, and C.~Moore, ``Modeling of ionized
  gas flows with a velocity-space hybrid {Boltzmann} solver,'' in {\em AIAA
  Scitech 2021 Forum}, p.~0705, 2021.

\bibitem{kolobov2007unified}
V.~Kolobov, R.~Arslanbekov, V.~V. Aristov, A.~Frolova, and S.~A. Zabelok,
  ``Unified solver for rarefied and continuum flows with adaptive mesh and
  algorithm refinement,'' {\em Journal of Computational Physics}, vol.~223,
  no.~2, pp.~589--608, 2007.

\bibitem{hauck2022predictor}
C.~Hauck and S.~Schnake, ``A predictor-corrector strategy for adaptivity in
  dynamical low-rank approximations,'' {\em arXiv preprint arXiv:2209.00550},
  2022.

\bibitem{einkemmer2021mass}
L.~Einkemmer and I.~Joseph, ``A mass, momentum, and energy conservative
  dynamical low-rank scheme for the {Vlasov} equation,'' {\em Journal of
  Computational Physics}, vol.~443, p.~110495, 2021.

\bibitem{guo2022local}
W.~Guo and J.-M. Qiu, ``A {Local Macroscopic Conservative (LoMaC)} low rank
  tensor method for the {Vlasov} dynamics,'' {\em arXiv preprint
  arXiv:2207.00518}, 2022.

\bibitem{einkemmer2021efficient}
L.~Einkemmer, J.~Hu, and L.~Ying, ``An efficient dynamical low-rank algorithm
  for the {Boltzmann-BGK} equation close to the compressible viscous flow
  regime,'' {\em SIAM Journal on Scientific Computing}, vol.~43, no.~5,
  pp.~B1057--B1080, 2021.

\bibitem{xiao2021using}
T.~Xiao and M.~Frank, ``Using neural networks to accelerate the solution of the
  {Boltzmann} equation,'' {\em Journal of Computational Physics}, vol.~443,
  p.~110521, 2021.

\bibitem{holloway2021acceleration}
I.~Holloway, A.~Wood, and A.~Alekseenko, ``Acceleration of {Boltzmann}
  collision integral calculation using machine learning,'' {\em Mathematics},
  vol.~9, no.~12, p.~1384, 2021.

\bibitem{chen2021unsupervised}
G.~Chen, L.~Chac{\'o}n, and T.~B. Nguyen, ``An unsupervised machine-learning
  checkpoint-restart algorithm using {Gaussian} mixtures for particle-in-cell
  simulations,'' {\em Journal of Computational Physics}, vol.~436, p.~110185,
  2021.

\bibitem{kornev2020tensorized}
E.~Kornev and A.~Chikitkin, ``A tensorized version of {LU-SGS} solver for
  discrete velocity method for {Boltzmann} kinetic equation with model
  collision integral,'' in {\em AIP Conference Proceedings}, vol.~2312,
  p.~050009, AIP Publishing LLC, 2020.

\bibitem{chikitkin2021numerical}
A.~V. Chikitkin, E.~Kornev, and V.~A. Titarev, ``Numerical solution of the
  {Boltzmann} equation with {S}-model collision integral using tensor
  decompositions,'' {\em Computer Physics Communications}, vol.~264, p.~107954,
  2021.

\bibitem{oseledets2011tensor}
I.~V. Oseledets, ``Tensor-train decomposition,'' {\em SIAM Journal on
  Scientific Computing}, vol.~33, no.~5, pp.~2295--2317, 2011.

\bibitem{khoromskij2007structured}
B.~Khoromskij, ``Structured data-sparse approximation to high order tensors
  arising from the deterministic {Boltzmann} equation,'' {\em Mathematics of
  computation}, vol.~76, no.~259, pp.~1291--1315, 2007.

\bibitem{dolgov2014low}
S.~V. Dolgov, A.~P. Smirnov, and E.~Tyrtyshnikov, ``Low-rank approximation in
  the numerical modeling of the {Farley--Buneman} instability in ionospheric
  plasma,'' {\em Journal of Computational Physics}, vol.~263, pp.~268--282,
  2014.

\bibitem{allmann2022parallel}
F.~Allmann-Rahn, R.~Grauer, and K.~Kormann, ``A parallel low-rank solver for
  the six-dimensional {Vlasov}--{Maxwell} equations,'' {\em Journal of
  Computational Physics}, vol.~469, p.~111562, 2022.

\bibitem{hu2022adaptive}
J.~Hu and Y.~Wang, ``An adaptive dynamical low rank method for the nonlinear
  {Boltzmann} equation,'' {\em Journal of Scientific Computing}, vol.~92,
  no.~2, p.~75, 2022.

\bibitem{mouhot2006fast}
C.~Mouhot and L.~Pareschi, ``Fast algorithms for computing the {Boltzmann}
  collision operator,'' {\em Mathematics of computation}, vol.~75, no.~256,
  pp.~1833--1852, 2006.

\bibitem{goldstein1989investigations}
D.~Goldstein, B.~Sturtevant, and J.~Broadwell, ``Investigations of the motion
  of discrete-velocity gases,'' {\em Progress in Astronautics and Aeronautics},
  vol.~117, pp.~100--117, 1989.

\bibitem{tcheremissine2005direct}
F.~Tcheremissine, ``Direct numerical solution of the {Boltzmann} equation,'' in
  {\em AIP Conference Proceedings}, vol.~762, pp.~677--685, American Institute
  of Physics, 2005.

\bibitem{varghese2007arbitrary}
P.~Varghese, ``Arbitrary post-collision velocities in a discrete velocity
  scheme for the {Boltzmann} equation,'' in {\em Proc. of the 25th Intern.
  Symposium on Rarefied Gas Dynamics/Ed. by MS Ivanov and AK Rebrov.
  Novosibirsk}, pp.~225--232, 2007.

\bibitem{morris2008improvement}
A.~Morris, P.~Varghese, and D.~Goldstein, ``Improvement of a discrete velocity
  {Boltzmann} equation solver with arbitrary post-collision velocities,'' in
  {\em AIP Conference Proceedings}, vol.~1084, pp.~458--463, American Institute
  of Physics, 2008.

\bibitem{poondla2020modeling}
Y.~K. Poondla, {\em Modeling reactive rarefied systems using a novel
  quasi-particle {Boltzmann} solver}.
\newblock PhD thesis, The University of Texas at Austin, 2020.

\bibitem{baranger2019numerical}
C.~Baranger, N.~H{\'e}rouard, J.~Mathiaud, and L.~Mieussens, ``Numerical
  boundary conditions in {Finite Volume} and {Discontinuous Galerkin} schemes
  for the simulation of rarefied flows along solid boundaries,'' {\em
  Mathematics and Computers in Simulation}, vol.~159, pp.~136--153, 2019.

\bibitem{zniyed2020tt}
Y.~Zniyed, R.~Boyer, A.~L. De~Almeida, and G.~Favier, ``A {TT}-based
  hierarchical framework for decomposing high-order tensors,'' {\em SIAM
  Journal on Scientific Computing}, vol.~42, no.~2, pp.~A822--A848, 2020.

\bibitem{boyd1996conservative}
I.~D. Boyd, ``Conservative species weighting scheme for the direct simulation
  monte carlo method,'' {\em Journal of Thermophysics and Heat Transfer},
  vol.~10, no.~4, pp.~579--585, 1996.

\bibitem{bezanson2017julia}
J.~Bezanson, A.~Edelman, S.~Karpinski, and V.~B. Shah, ``Julia: A fresh
  approach to numerical computing,'' {\em SIAM review}, vol.~59, no.~1,
  pp.~65--98, 2017.

\bibitem{kressner2017recompression}
D.~Kressner and L.~Perisa, ``Recompression of {Hadamard} products of tensors in
  {Tucker} format,'' {\em SIAM Journal on Scientific Computing}, vol.~39,
  no.~5, pp.~A1879--A1902, 2017.

\bibitem{ttjl}
A.~Periša, ``{TensorToolbox.jl}'', \url{https://doi.org/10.5281/zenodo.4627980},
  \url{https://github.com/lanaperisa/TensorToolbox.jl}, 2021.

\bibitem{bobylev1976}
A.~V. Bobylev, ``One class of invariant solutions of the \protect{Boltzmann}
  equation,'' in {\em Akademiia Nauk SSSR Doklady}, vol.~231, pp.~571--574,
  1976.

\bibitem{krook1977}
M.~Krook and T.~T. Wu, ``Exact solutions of the \protect{Boltzmann} equation,''
  {\em Phys. Fluids}, vol.~20, no.~10, pp.~1589--1595, 1977.

\end{thebibliography}

\end{document}